\documentclass[draftcls,12pt,onecolumn]{IEEEtran}
\usepackage{epsfig}
\usepackage{amsmath,amssymb}
\usepackage{color}
\usepackage{algorithm}
\usepackage{algorithmic}
\usepackage{multirow}
\usepackage{graphicx}
\usepackage{subfig}
\usepackage{cite}
\usepackage{longtable}
\usepackage{chngpage}
\usepackage{array}
\usepackage{booktabs}
\usepackage[printonlyused]{acronym}
\usepackage[table]{xcolor}

\usepackage[style=list,nonumberlist]{glossaries}

\newglossaryentry{2D}{name=2D, description={Two-Dimensional}}
\newglossaryentry{3D}{name=3D, description={Three-Dimensional}}
\newglossaryentry{4G}{name=4G, description={Fourth Generation}}
\newglossaryentry{5G}{name=5G, description={Fifth Generation}}
\newglossaryentry{3GPP}{name=3GPP, description={Third Generation Partnership Project}}

\newglossaryentry{B5G}{name=B5G, description={Beyond Fifth Generation}}
\newglossaryentry{BS}{name=BS, description={Base Station}}
\newglossaryentry{CR}{name=CR, description={Cognitive Radio}}
\newglossaryentry{D2D}{name=D2D, description={Device-to-Device}}
\newglossaryentry{eMBB}{name=eMBB, description={Enhanced Mobile Broadband}}
\newglossaryentry{FANET}{name=FANET, description={Flying Ad Hoc Network}}
\newglossaryentry{GCS}{name=GCS, description={Ground Control Station}}

\newglossaryentry{HAP}{name=HAP, description={High Altitude Platform}}
\newglossaryentry{HetNet}{name=HetNet, description={Heterogeneous Network}}
\newglossaryentry{IoT}{name=IoT, description={Internet of Things}}
\newglossaryentry{IoUAVs}{name=IoUAVs, description={Internet of UAVs}}
\newglossaryentry{LAP}{name=LAP, description={Low Altitude Platform}}
\newglossaryentry{LoS}{name=LoS, description={Line-of-Sight}}
\newglossaryentry{MBS}{name=MBS, description={Macro Base Station}}
\newglossaryentry{MBSs}{name=MBSs, description={Macro Base Stations}}
\newglossaryentry{MEC}{name=MEC, description={Mobile Edge Computing}}
\newglossaryentry{MIMO}{name=MIMO, description={Multiple-Input Multiple-Output}}
\newglossaryentry{mMTC}{name=mMTC, description={Massive Machine-Type Communications}}
\newglossaryentry{mmWave}{name=mmWave, description={Millimeter-Wave}}
\newglossaryentry{NFV}{name=NFV, description={Network Functions Virtualization}}
\newglossaryentry{NOMA}{name=NOMA, description={Non-Orthogonal Multiple Access}}

\newglossaryentry{OMA}{name=OMA, description={Orthogonal Multiple Access}}
\newglossaryentry{QoE}{name=QoE, description={Quality of Experience}}
\newglossaryentry{QoS}{name=QoS, description={Quality of Service}}

\newglossaryentry{RF}{name=RF, description={Radio Frequency}}
\newglossaryentry{SBS}{name=SBS, description={Small Base Station}}
\newglossaryentry{SBSs}{name=SBSs, description={Small Base Stations}}
\newglossaryentry{SDN}{name=SDN, description={Software-Defined Networking}}
\newglossaryentry{SIC}{name=SIC, description={Successive Interference Cancellation}}
\newglossaryentry{SIMO}{name=SIMO, description={Single-Input Multiple-Output}}
\newglossaryentry{SINR}{name=SINR, description={Signal-to-Interference-plus-Noise Ratio}}
\newglossaryentry{SIR}{name=SIR, description={Signal-to-Interference Ratio}}
\newglossaryentry{SNR}{name=SNR, description={Signal-to-Noise Ratio}}
\newglossaryentry{SWAP}{name=SWAP, description={Size, Weight and Power}}

\newglossaryentry{UAV}{name=UAV, description={Unmanned Aerial Vehicle}}
\newglossaryentry{URLLC}{name=URLLC, description={Ultra-Reliable Low-Latency Communication}}
\newglossaryentry{SWIPT}{name=SWIPT, description={Simultaneous Wireless Information and Power Transfer}}
\newglossaryentry{WPCN}{name=WPCN, description={Wireless Powered Communication Network}}
\newglossaryentry{WPT}{name=WPT, description={Wireless Power Transfer}}

\newglossaryentry{DF}{name=DF, description={decode-and-forward}}

\newglossaryentry{ICI}{name=ICI, description={interchannel interference}}
\newglossaryentry{ICT}{name=ICT, description={information and communication technology}}

\newglossaryentry{LTE}{name=LTE, description={long-term evolution}}
\newglossaryentry{LTE-A}{name=LTE-A, description={Long Term Evolution-Advanced}}

\newglossaryentry{TDD}{name=TDD, description={time-division-duplex}}
\newglossaryentry{TDM}{name=TDM, description={time-division multiplexing}}
\newglossaryentry{TDMA}{name=TDMA, description={time-division multiple-access}}

\newglossaryentry{WBAN}{name=WBAN, description={wireless body area network}}

\newglossaryentry{WLAN}{name=WLAN, description={wireless local area network}}

\makeglossaries

\begin{document}
\title{UAV Communications for 5G and Beyond:  Recent Advances and Future Trends}
\author{Bin Li, Zesong Fei,~\IEEEmembership{Senior Member,~IEEE}, and Yan Zhang,~\IEEEmembership{Senior Member,~IEEE}
\thanks{This work was supported in part by the National Natural Science Foundation of China under Grant 61871032, in part by
the China National S\&T Major Project under Grant 2017ZX03001017, in part by 111 Project of China under Grant B14010, in part by the projects 240079/F20 funded by the Research Council of Norway, and in part by the scholarship from China Scholarship Council [2017] 3019. (\textit{Corresponding author: Zesong Fei.})}

\thanks{B. Li is with the School of Information and Electronics, Beijing Institute of Technology, Beijing 100081, China, and also with the School of Computer and Software, Nanjing University of Information Science and Technology, Nanjing 210044, China
(e-mail: libin\_sun@bit.edu.cn).}
\thanks{Z. Fei is with the School of Information and Electronics, Beijing Institute of Technology, Beijing 100081, China (e-mail: feizesong@bit.edu.cn).}
\thanks{Y. Zhang is with the Department of Informatics, University of Oslo, Norway, and also with the Simula Research Laboratory, Norway (e-mail:
yanzhang@ieee.org).}
}

\markboth{{This} {article} {has} {been} {accepted} {for} {publication} {in} {a} {future} {issue} {of} IEEE Internet of Things Journal.}
{Shell \MakeLowercase{\textit{et al.}}: Bare Demo of IEEEtran.cls for Journals}

\maketitle

\begin{abstract}
Providing ubiquitous connectivity to diverse device types is the key challenge for 5G and beyond 5G (B5G). Unmanned aerial vehicles (UAVs) are expected to be an important component of the upcoming  wireless networks that can potentially facilitate wireless broadcast and support high rate transmissions.
Compared to the communications with fixed infrastructure, UAV has salient attributes, such as  flexible deployment, strong line-of-sight (LoS) connection links, and additional design degrees of freedom with the controlled mobility. In this paper, a comprehensive survey on UAV communication towards 5G/B5G wireless networks is presented. We first briefly introduce essential background and the space-air-ground integrated networks, as well as discuss related research challenges  faced by the emerging  integrated network architecture. We then provide an exhaustive review of various 5G techniques based on UAV platforms, which we categorize by different domains including physical layer, network layer, and joint communication, computing and caching.
In addition, a great number of open research problems are outlined and identified as possible future research directions.

\end{abstract}


\IEEEpeerreviewmaketitle
{\footnotesize{\printglossaries}}

\section{Introduction \label{a}}

\IEEEPARstart{T}{he} landscape of future fifth generation (\gls{5G}) radio access networks is expected to seamlessly and ubiquitously connect everything, and support at least 1000-fold traffic volumes, 100 billion connected wireless devices, and diversified requirements on reliability,
latency, battery lifetime, etc, as opposed to current fourth generation (\gls{4G}) cellular networks. Nowadays, the popularity of the Internet of Things (\gls{IoT}) has triggered a surge in the number of mobile data traffic for upcoming 5G and beyond 5G (\gls{B5G}) wireless networks. In accordance with the latest report \cite{Khan2016multi}, the global mobile traffic will reach $1$ zettabyte/mo until $2028$. This will lead the current infrastructure facing great capacity demands and also impose a heavy burden on the telecom operators in terms of increased capital investments and operational costs.
Some early efforts have been dedicated to heterogeneous networks (\glspl{HetNet}) (i.e., deploy various small cells) to meet these growing demands \cite{Zhang2014Enhancing}.

However, in unexpected or emergency situations (such as disaster relief and service recovery), the deployment of terrestrial
infrastructures is economically infeasible and challenging due to high operational expenditure as well as sophisticated and volatile environments.
To handle this issue, intelligent heterogenous architecture by leveraging unmanned aerial vehicles (\glspl{UAV}) (or commonly known as drones) \cite{Zeng2016Mag} has been considered to be a promising new paradigm to facilitate three central usage scenarios of future wireless networks, i.e., enhanced mobile broadband (\gls{eMBB}) with bandwidth-consuming, ultra-reliable low-latency communication (\gls{URLLC}) and massive machine-type
communications (\gls{mMTC}). For instance, UAV may play a central role in providing network service recovery in a disaster-stricken region, enhancing public safety networks, or handling other emergency situations when URLLC is required. In particular, UAV-aided eMBB can be regarded as an important complement to the 5G cellular networks \cite{Huo2018Distributed}. As a result, UAVs are identified as an important component
of 5G/B5G wireless technologies.

Owing to the versatility and high mobility of UAVs, low-altitude UAVs are extensively used in diverse fields for different applications and purposes.
On the standpoint of  wireless  communication aspects,
UAVs can be employed as aerial communication platforms (e.g., flying base stations (\glspl{BS}) or mobile relays) by mounting communication transceivers to provide/enhance communication services to ground targets in high traffic demand and overloaded situations,  which is  commonly referred to as \textit{UAV-assisted communications} \cite{Mozaffari2017TWC,Zeng2016TCom,Mozaffari2017Mobile,Li2018GLOBECOMSecrecy,Qi2019SEE}. On the other hand, UAVs can also
be used as aerial nodes to enable a multitude of applications ranging from cargo delivery to surveillance, which is  commonly referred to as \textit{cellular-connected UAVs} \cite{Zeng2018cellular,Mozaffari2018Beyond}. Most of the existing body of work, however, is restricted to UAVs in the role of
assisting cellular communications.  In most current contexts, UAVs are equipped with communication devices or dedicated sensors  that can enable a myriad of applications such as low altitude surveillance, post-disaster rescue, logistics application and communication assistance. Furthermore, to support broadband wireless communications in a large geographical area, a swarm of UAVs forming Flying Ad Hoc Networks (\glspl{FANET}) \cite{Hayat2016survey,Sekander2018multi} and establishing connection links with the ground nodes, have been studied theoretically and validated through field experiments.
As a desirable candidate to substitute or complement terrestrial cellular networks, UAV communications exhibit major attributes as follows \cite{Shi2018drone}:
\begin{itemize}
\item Line-of-Sight Links: UAVs without human pilots flying in the sky have a higher probability to connect ground users via line-of-sight (\gls{LoS}) links, which facilitates highly reliable transmissions over long distances. In addition, UAVs can adjust their hovering locations to maintain the quality of links.

\item Dynamic Deployment Ability:
Compared with stationary ground infrastructures, UAVs can be dynamically deployed according to real-time
requirements, which is more robust against the environment changes. In addition, UAVs as aerial BSs do not require the site rental costs, thus removing the need for towers and cables.

\item UAV-Based Swarm Networks: A swarm of UAVs are capable of forming scalable multi-UAV networks and offering ubiquitous connectivity
 to ground users. Benefiting from its high flexibility and rapid provision features, the multi-UAV network is a feasible solution to recover and expand
communication in fast and effective ways.
\end{itemize}

In fact, UAVs are distinguished according to the stringent constraint imposed by the size, weight, and power  (\gls{SWAP}), since the SWAP constraint directly impacts the maximum operational altitude, communication, coverage, computation, and endurance capabilities of each UAV. For instance, low altitude platforms
(\glspl{LAP}) have low power and low capacity in terms of both payload and autonomy. By contrast, higher altitude platforms (\glspl{HAP}) provide wider coverage and longer endurance \cite{Mozaffari2018tutorial}. As the altitude of UAV increases, the probability of having an LoS link for air-to-ground communication increases, mainly due to a higher
probability of observing an unobstructed path. Meanwhile, the path loss is more severe due to the increased distance
between the UAV and ground users. Thus, the two opposing aspects on the UAV's altitude need a
fundamental trade-off while guaranteeing the maximum cell coverage.

It is noteworthy that 5G/B5G wireless networks are expected to exhibit great heterogeneities in communication infrastructures and resources for connecting different devices and providing diverse services \cite{Naqvi2018drone}. Researchers these days are focusing on ways to design heterogeneous infrastructures such as densely deployed  small cells;
integrate heterogeneous communication networks such as space-based, air-based, and ground-based \cite{Zhang2017software};
employ multifarious 5G communication techniques \cite{Agiwal2016next} such as massive multiple-input multiple-output (MIMO), millimeter-wave (\gls{mmWave}), non-orthogonal multiple access (\gls{NOMA}) transmission, device-to-device (\gls{D2D}), cognitive radio (\gls{CR}) and so forth, to improve spectrum efficiency and energy efficiency.
Regarding the UAV-assisted cellular networks, the operation cost (e.g., endurance time) is one of the most important factors. For this reason, energy harvesting can be a  must-has core technology.
In a meantime, UAVs can serve as edge network controllers to efficiently allocate computing and storage resources. Particularly, UAVs can either
serve as edge computing platforms to offload the computing tasks from IoT devices, or cache popular contents to reduce the burden of backhaul networks  \cite{Cheng2018air}.

\subsection{Existing Surveys and Tutorials}

A couple of surveys and tutorials related to UAV communications have been published over the past several years \cite{Gupta2016survey,Hayat2016survey,Motlagh2016low,Krishna2017review,Jiang2018routing,Khawaja2018survey,Khuwaja2018survey,Lu2018Wireless,Cao2018Airborne,Mozaffari2018tutorial},
including the characteristics and requirements of UAV networks, main communication issues, cyber-security, wireless charging
techniques, and channel modeling for UAV communications, etc.

More specifically, Hayat \textit{et al.} \cite{Hayat2016survey} have reviewed the civil applications of UAV networks from a communication  perspective along with its characteristics. They also reported experimental results from many projects.
A survey paper by Gupta \textit{et al.} \cite{Gupta2016survey} elaborated many issues encountered in UAV communication networks to provide stable and reliable wireless transmission. Motlagh \textit{et al.} \cite{Motlagh2016low} presented a comprehensive survey and highlighted the potentials for the delivery of low altitude UAV-based IoT services from the sky.  The cybersecurity for UAVs was reviewed in \cite{Krishna2017review}, where actual and simulated attacks were discussed. Furthermore, Jiang \textit{et al.} \cite{Jiang2018routing}  surveyed the most representative  routing protocols for UAVs and compared the performance of the existing major routing protocols.
Another survey by Khawaja \textit{et al.} \cite{Khawaja2018survey} solely focused on the  air-to-ground propagation channel measurement and modeling. They also discussed various channel characterization efforts.
While in \cite{Khuwaja2018survey}, from a channel modeling viewpoint, Khuwaja \textit{et al.}
reported the extensive measurement methods for UAV channel modeling based on the LAPs and discussed various channel characteristics.
Lu \textit{et al.} \cite{Lu2018Wireless}  introduced various prevalent wireless charging techniques conceived for UAV mission time improvement.
They provided a classification of wireless charging techniques, namely the family of non-electromagnetic-based and the family of electromagnetic-based methods.
Cao \textit{et al.} \cite{Cao2018Airborne} were concerned with the
primary mechanisms and protocols for the design of airborne communication networks by considering the LAP-based communication networks, the HAP-based communication networks, and the integrated communication networks.
Additionally, in a more recent study Mozaffari \textit{et al.} \cite{Mozaffari2018tutorial} provided
a holistic tutorial on UAV-enabled wireless networks and reviewed various analytical frameworks and mathematical tools conceived for solving fundamental open problems.
The above-mentioned surveys related to UAV communications are outlined at a glance in Table \ref{relatedworks}, which
allows the readers to capture the main contributions of each of the existing surveys.

\begin{table*}\small
\centering
\setlength{\tabcolsep}{2pt}
\renewcommand{\arraystretch}{1.3}
\extrarowheight 3pt
\caption{Existing Surveys Relating to UAV Communications.}
\begin{tabular}{| l | c | }
\rowcolor{red!15}
\hline
\textbf{Publication} & \textbf{One-sentence summary}  \\
\hline
\hline
Hayat \textit{et al.} \cite{Hayat2016survey} & A survey of the characteristics and requirements of UAV networks  \\
\hline
Gupta \textit{et al.} \cite{Gupta2016survey} & A survey on the main issues in UAV communications networks  \\
\hline
Motlagh \textit{et al.} \cite{Motlagh2016low} & A comprehensive survey on UAVs-based IoT services  \\
\hline
Krishna \textit{et al.} \cite{Krishna2017review}  & A review on cybersecurity for UAVs   \\
\hline
Jiang \textit{et al.} \cite{Jiang2018routing}  & A survey of routing protocols for UAVs \\
\hline
Khawaja \textit{et al.} \cite{Khawaja2018survey}  & An overview of air-to-ground propagation channel modeling \\
\hline
Khuwaja \textit{et al.} \cite{Khuwaja2018survey}  & A survey of the measurement methods proposed for UAV channel modeling \\
\hline
Lu \textit{et al.} \cite{Lu2018Wireless}  & Review of wireless charging techniques for UAVs \\
\hline
Cao \textit{et al.} \cite{Cao2018Airborne}  & Overview of airborne communication networks \\
\hline
Mozaffari \textit{et al.} \cite{Mozaffari2018tutorial} & A comprehensive tutorial on the use of UAVs in wireless networks  \\
\hline
\end{tabular}
\label{relatedworks}
\end{table*}

\subsection{Paper Contributions and Organization}
Although the aforementioned existing studies provided insights into several perspectives for UAV communication networks, it is worth reflecting upon the current achievements in order to shed light on the future research trends for 5G/B5G.
Therefore, it is of great importance and necessity to provide an overview of the emerging studies related to the integration of 5G technologies with UAV communication networks. In this survey,
we are intending to provide the reader with an emerging space-air-ground integrated network architecture and highlight a variety of open research challenges. Then, we present an exhaustive review of the up-to-date
research progress of UAV communications integration with various 5G technologies at (i) physical layer,
(ii) network layer, and (iii) joint communication, computing, and caching. Finally, we identify possible future trends for UAV communications according to the latest developments.

\begin{figure}[t]
\centering
\includegraphics[width=4.0in]{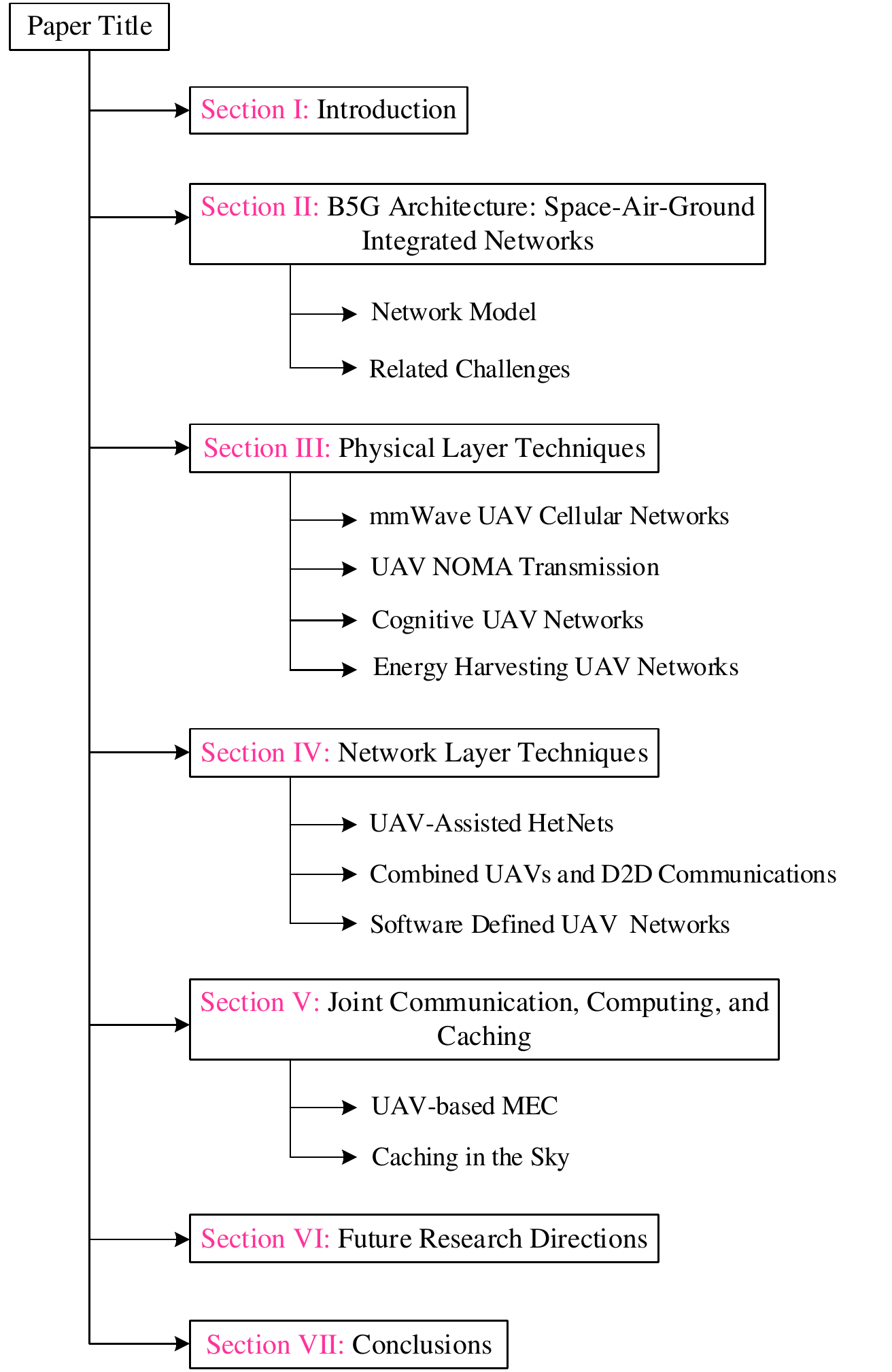}
\caption{The organization of this paper.}
\label{organize}
\end{figure}

The rest of this article is outlined as follows. In Section
\ref{b}, we envision an overview of the space-air-ground integrated network and discuss the relevant challenges for the emerging architecture. In Section \ref{c}, we provide a physical layer overview of the state-of-the-art studies dedicated to integrating UAV into 5G/B5G communications.
In Section \ref{d}, we present a network layer overview of the existing studies dedicated to integrating UAV into 5G/B5G communications.
In Section \ref{e}, we review the existing contributions on joint communication, computing, and caching for UAV communications.
Finally, in Section \ref{f} we describe a range of open problems to be tackled by future research, followed by our conclusions in Section \ref{g}.
For the sake of explicit clarity, the organization of this paper is shown in Fig. \ref{organize}.

\section{B5G Architecture: Space-Air-Ground Integrated Networks \label{b}}
In this section, we first present the space-air-ground integrated network architecture in upcoming 5G/B5G wireless communications,  where a three-layer cooperative network is introduced and explained briefly. Then, we discuss the major challenges faced by the system design.

\subsection{Space-Air-Ground Integrated Networks}

\begin{figure*}
\centering
\includegraphics[width=5.0in]{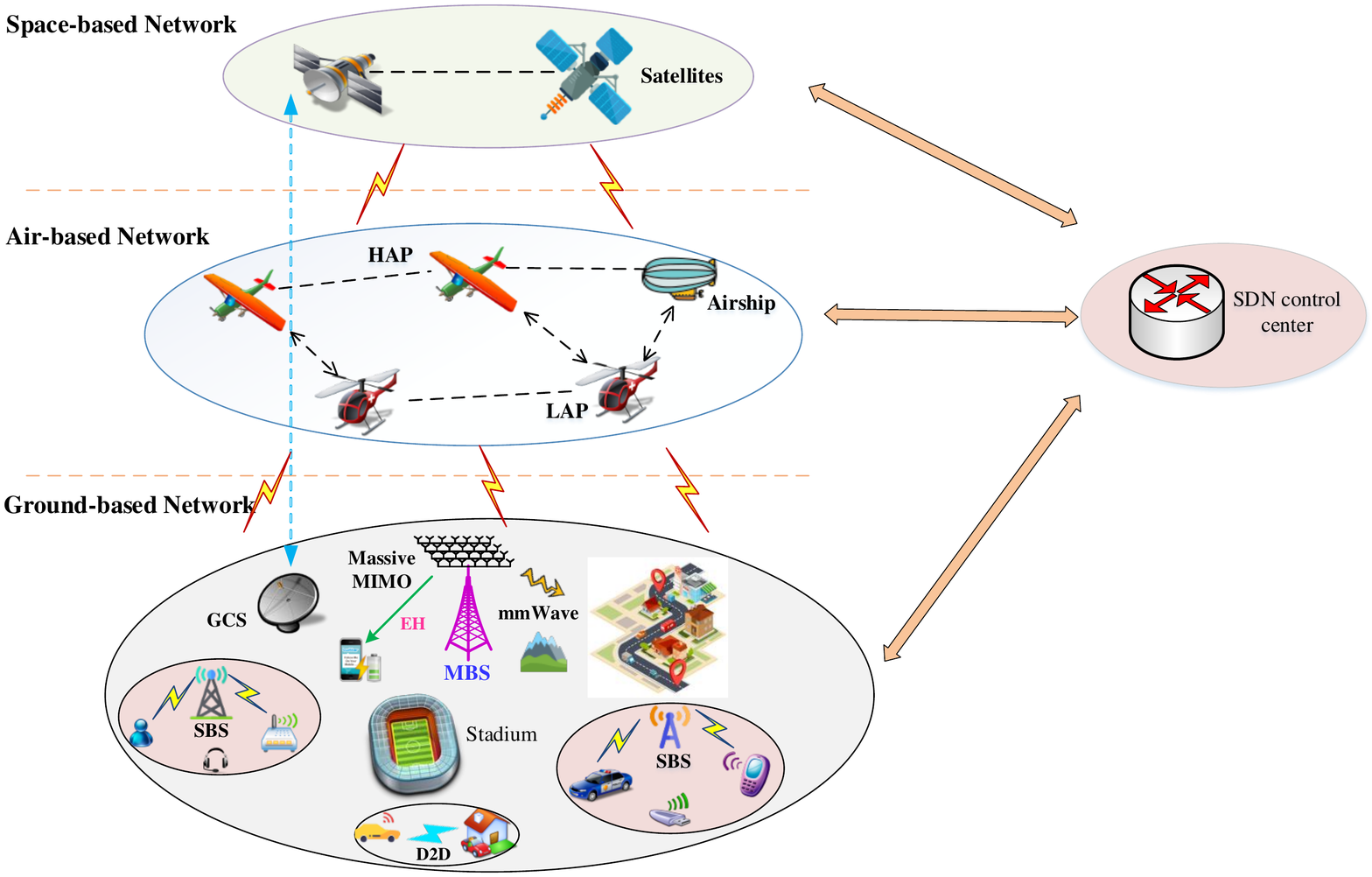}
\caption{Illustration of space-air-ground integrated networks.  The conceptual communication architecture in UAV-aided networks is composed of three layers: the space layer for the satellite data links, the air layer for UAVs and LoS data links, and the ground layer for end user devices, GCS, government/security center.
}
\label{sys}
\end{figure*}

To accommodate the diverse IoT services with different quality
of service (\gls{QoS}) requirements in various practical scenarios (e.g.,
urban, rural and sparsely populated areas) \cite{Zhang2011Home}, it is imperative to exploit specific advantages of each networking paradigm.
For instance, densely deployed terrestrial networks in urban areas can support high data rate access, satellite communication systems can provide wide coverage and seamless connectivity to the most remote and sparsely populated areas, while
UAV communications can assist the existing cellular communications for the rapid service recovery and offer the traffic offloading of the extremely crowded areas in a cost-effective fashion \cite{Liu2017energy}. At present, it is widely believed that the individually existing network cannot meet the need to process enormous volumes of data and execute substantial applications such as IoT, cloud computing, and big data. Therefore, there is a
growing demand among scientific communities to develop an integrated network architecture from the space-based network, air-based network and ground-based network.

The overall architecture of the space-air-ground integrated network is presented in Fig. \ref{sys} to provide user devices with improved and flexible end to end
services, which is categorized into three segments, i.e., space-based, air-based, and ground-based layers. Thereinto, UAVs are deployed to set up a multi-tier UAV network, as well as the radio access infrastructure, the mobile users and vehicles form the ground network.
At the same time, software defined networking (\gls{SDN})  controllers \cite{Zhang2017software} can be deployed to
regulate the network behaviors and manage the network resources in an agile and flexible manner (in software) to facilitate the space-air-ground interworking. Considering different segments having distinct characteristics, such as communication standards and diverse network devices with various functions, the control and communication interfaces of SDN controllers for each segment should be dedicated to the corresponding segment.

Regarding the space-based network, it is composed of a number of satellites or constellations of different orbits (like geostationary earth orbit (35786 km), low earth orbit (700-2000 km) and medium earth orbit (8000-20000 km)), ground stations, and network operations control centers. Satellites in different orbits, types and properties can form a global space-based network through inter-satellite links, which in turn utilize the multi-cast and broadcast techniques to improve the network capacity.
Meanwhile, by establishing the satellite-to-UAV and satellite-to-ground links, the connections are created with their neighbor satellites and ground cellular networks. On the basis, the space-based network can provide global coverage on the earth with services for emergency rescue, navigation, earth observation, and communication/relaying.
We can imagine that the future earth will be surrounded by large volumes of satellites.
However, the data delivery between the satellites and the ground segment is affected by long transmission latencies in virtue of the large free-space path loss and tropospheric attenuation. It is compelling to use higher frequency band for providing low-latency and high-throughput services, such as C-band and Ka-band \cite{Liujiajia2018survey}.

Satellite-to-UAV communication is a key component for building the integrated space-air-ground network. It is worth noting that the satellite-to-UAV channel mainly relies on the LoS link and also suffers from the rain attenuation significantly when using the Ka-band and above. In light of its applications and equipment, UAV can communicate with the satellites in different orbits during UAV navigation.
Generally, geosynchronous satellite is used for satellite-to-UAV communication since its location relative to earth keeps invariant \cite{Zhao2018JSACBeam}.
For UAV-to-satellite link, the premise of establishing a successful link is the alignment of the spatial beam from UAV to target satellite (i.e., the direction). Nevertheless, the continuous navigation of UAVs would result in the attitude variation all the time, which directly affects the spatial beam pointing for the UAV-to-satellite link. One typical scenario is UAV-assisted
satellite communication, where UAV needs to constantly adjust its beam towards the target satellite to maintain the communication link \cite{Zhao2018AccessIntegrating}.

In the air-based network,  a wide variety of unmanned flying platforms including UAVs, airships or balloons may be restricted to different operational altitudes due to the SWAP constraints.  Generally, a UAV is equipped with  transceivers to  provide flexible Internet access for a group of ground users and a \textit{drone-cell} is the corresponding coverage area. The size of drone-cell is dominated by UAV's altitude, location, transmission power, and the environment factors.
Furthermore, a swarm of UAVs are connected by establishing the UAV-to-UAV links to provide services cooperatively.
The multi-tier UAV network not only supports control messages exchanging among UAVs to avoid collisions and calculate flying paths, but also sends data to mobile devices accessing them. Specific UAVs are outfitted with heterogeneous radio interfaces, such as
LTE or WiFi, to communicate with infrastructures or satellites, which establishes gateways between multi-tier UAV networks and other networks.
The UAV can either use a sky-haul link to the satellites or connect to the ground system via a backhaul link \cite{Alzenad2018fso}.

In the ground-based network, the heterogeneous  radio access network comprising of macro cells and small cells serves the mobile users, such as mobile phones, self-driving cars, IoT devices, and so forth,  which will create an coexistent system of disruptive technologies for 5G wireless networks.
This covers all promising 5G cellular technologies, including mmWave frequency band, energy harvesting, NOMA transmission, and D2D communication,  as shown in Fig. \ref{sys}, and has become an important research topic recently.
In addition, the exponentially increasing computing capability of mobile devices can be conceived for mobile edge computing (\gls{MEC}), where UAVs can schedule the computing tasks while onboard computers fulfill these tasks.
And along with this, popular contents can be cached at the UAVs or ground devices, and transmitted through the drone-cells or D2D communication between end devices.
In particular, there are two kinds of transmission channels in the integration of air-based network and ground-based network: the LoS data link and the satellite data link. The LoS data link is used for direct transmission from the UAV to the ground control station (\gls{GCS}), in which light emissions travel in a straight line. In this kind of transmission, the waves may be easily absorbed by obstructions, so it is not suitable for the military fields.

In this paper, a general B5G integrated network architecture has been proposed. It describes the interconnectivity
among the different emerging technologies. The concept of mmWave frequency band, energy harvesting, NOMA transmission, and D2D communication has also been incorporated in this proposed B5G network architecture.
This proposed architecture also explains the roles of MEC and cache. In general, the proposed integrated network architecture may provide a good platform for future network standardization, which is expected to be reliable, real-time, efficient and safety.

\subsection{Potential Challenges}

Although the importance of the space-air-ground integrated network in B5G wireless communications is increasingly growing, developing the integrated network is a challenging task \cite{Bor2016new} that includes air-to-ground channel modeling, optimal deployment, energy efficiency, path planning, resource management, network security, etc.
In this subsection, we summarize in detail the key challenges faced by future space-air-ground integrated networks as follows:

\textbf{Channel Modeling:}
Due to the distinctive channel characteristics of the air segment (such as three-dimensional (\gls{3D}) space and time-variability), the UAV-to-ground channels are
much more complex than current ground communication channels \cite{Yaliniz2018Environment}. Also, the UAV-to-ground channels are more susceptible
to blockage than the air-to-air communication links that experience dominant LoS. Therefore, the conventional models are often not well suited
for characterizing UAV-to-ground channels.
Given the heterogeneous environment, the UAV-to-ground channels are highly dependent
on the altitude and type of the UAV, the elevation angle, and the type of the propagation environment.
Clearly, accurate channel modeling link is of vital importance to evaluate the system performance properly.
However, finding a generic channel model for UAV-to-ground communications in consideration of such factors is challenging, which needs comprehensive simulations and measurements in various environments.
Currently, many channel measurement campaigns and modeling efforts have been made to characterize the UAV-to-ground channels \cite{Yaliniz2016Efficient,Mozaffari2016Efficient,Ye2017AirG}.

\textbf{Deployment:}
UAV-satellite communication is a key component for building the integrated space-air-ground network,
the mobility of UAV and satellite complicates integrated network operation. On one hand, the characteristics of the air-to-ground channels need to be considered for optimal 3D deployment of UAVs to decrease handover and avoid physical collisions. On the other hand, a satellite system is limited in power and bandwidth suffering from large transmission delay, and the satellite-to-ground channel fading at high frequencies (typically Ka-band) is much more severe, which seemingly blocks it from practical usage.

\textbf{Path Planning:}
For an air-based network comprising the swarm of UAVs, each of which has a trajectory flying over the ground.
In order to reduce the communication delay, a UAV needs to
move close to the ground users. However, due to the need of keeping interconnection with its neighboring UAVs, it is not always
possible for a given UAV to maintain a close link with the served users.
As a consequence, if a swarm of UAVs are considered, finding an optimal flying trajectory for a UAV is an arduous  task due to practical constraints.
Thus, it is urgently needed to exploit a dynamic trajectory control method for UAVs to increase the probability of end-to-end link connections
while maintaining sufficient coverage of the entire target area.

\textbf{Operational Altitude:}
Due to SWAP constraints, different types of UAVs may be restricted to different operational altitudes. For instance, mobile devices in
urban scenarios may require higher LoS connectivity, whereas mobile devices in suburban scenarios
may need higher degree of path loss reduction. Note that the higher altitude of UAVs promotes
higher LoS connectivity since reflection and shadowing are diminished, while lower altitude
ensures reduction in path loss. By selecting different heights for multi-tier UAVs, an optimal trade-off
between LoS connectivity and path loss can be struck.

\textbf{Interference Dynamics:}
In the constructed multi-tier HetNets, the ground cellular network and the air-to-ground channel suffer from high co-channel interference from the same and different segments, which will gradually render the current air interface obsolete. Furthermore, the UAV's mobility creates Doppler shift, which also causes severe inter-carrier interference at high frequencies.
Hence, in consideration of mobile characteristics, appropriate interference management in the integrated network becomes more challenging.

\textbf{Limited Energy:}
Since UAVs mainly rely on rechargeable battery power, the cruising duration on UAVs is strongly affected by the energy consumption of UAVs which may depend on their mobility, transmission power, and circuit power consumption. This is a prime challenge that significantly limits their operation time. Naturally, it is crucial to prolong the service duration or even provide persistent service during the
mission via advanced charging technologies.

\textbf{Backhaul Cellular Communication:}
An important distinction between ground BSs and UAV-BSs is the fact that the backhaul network is characterized by heterogeneous links.
Specifically, ground BSs are typically connected with the core
network via wired links that have large bandwidth. By contrast, UAV-BSs connecting with the macro base stations (\glspl{MBS}) or the
core network need high-capacity wireless backhaul links. Practically, the limited backhauls will become the bottleneck and affect the QoS of mobile users.

\textbf{Network Security:}
As the integrated network creates a multi-tier topology where multiple nodes are deployed with dissimilar characteristics and the broadcasting nature of wireless LoS propagation, the integrated network is  particularly  vulnerable to malicious attacks. As a result, safeguard strategies or protocols are of paramount importance. Furthermore, the accurate positioning of the UAVs and the detection of unauthorized intrusion into the airspace is another open aspect. Besides, SDN controllers are mainly responsible for managing resources and controlling network operation,
protecting the SDN controllers from different cyber attacks is still a challenge in integrated networks.

\textbf{Real-Time Demand:} Satellites for different tasks or services have different velocities and communication coverage, the data links between nodes may be frequent intermittence owing to the high bit error rate and transmission latency. Since the satellite data link adopts an onboard satellite transmission system to transmit remote data, a challenge is how to maximize the fast data acquisition ability and real-time exchange of information capability while transmitting data to the GCS.

In the next three sections, we provide an overview of the existing contributions related to UAV communications in the context of 5G/B5G wireless networks. We review these state-of-the-art studies mainly based on the current 5G technologies from physical layer, network layer, as well as joint communication, computing and caching perspectives, which are intended to provide useful guidelines for researchers to understand the referenced literature.

\section{Physical Layer Techniques \label{c}}

\begin{figure*}
\centering
\includegraphics[width=5.0in]{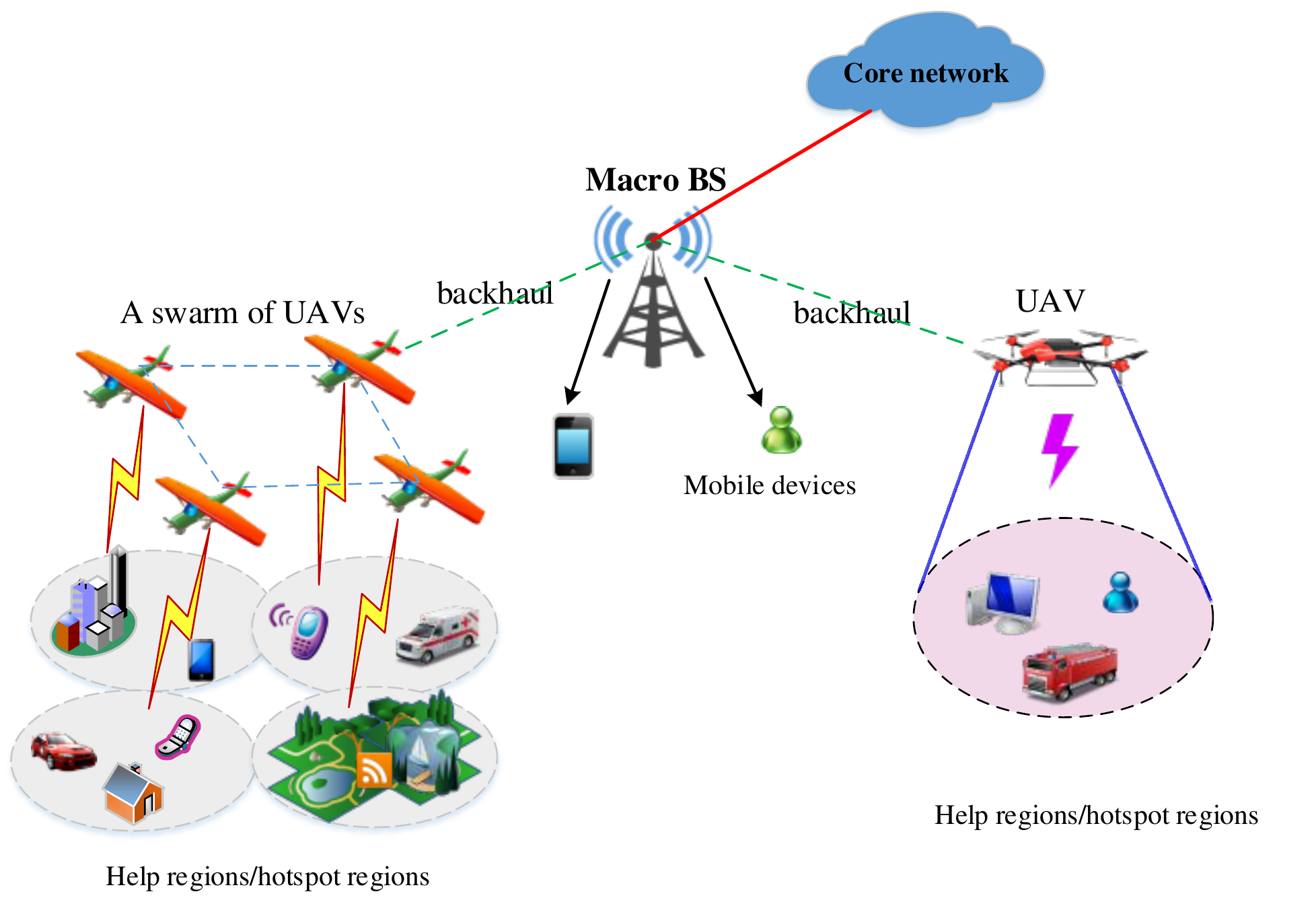}
\caption{An exemplary scenario of UAVs as aerial BSs serving a target area, where each UAV is equipped with wireless transceivers allowing them to communicate with
ground users and also with other UAVs.}
\label{UAVasBSs}
\end{figure*}

Currently, a variety of works are mainly concerned with the UAV-assisted communication networks,
especially in unexpected or temporary events \cite{Zhou2018Uplink}. Benefiting from the portable transceiver functionality and advanced signal processing techniques, the success of UAV communications can realize omnipresent coverage and support massive dynamic connections. Fig. \ref{UAVasBSs} depicts
the scenario of UAVs acting as flying BSs (i.e., UAV-BSs) where these UAVs are usually equipped with
diverse payloads for receiving, processing and transmitting signals, aiming to complement pre-existing cellular systems by providing additional capacity to
hotspot areas during temporary events. This scenario has been considered as one of the five key scenarios faced by future cellular networks \cite{Osseiran2014Mag}. Also, UAVs can be used to reinforce the communication infrastructure in emergency and
public safety situations during which the existing terrestrial network is damaged or not fully operational \cite{Yaliniz2018Strategic}.

In order to improve the system performance of UAV communication in 5G networks, physical layer techniques are of much concern as
they affect the applications of UAVs significantly. There are mainly five candidate key technologies at physical layer, namely mmWave communication, NOMA transmission, CR, and energy harvesting.
In this section, we review the state-of-the-art works on mmWave UAV-assisted cellular networks, UAV NOMA transmission, cognitive UAV networks, and energy harvesting UAV networks.

\subsection{{mm}Wave UAV-Assisted Cellular Networks}

It is crucial to note that UAVs may have to deal with different types of data such as voice, video and huge data files, which creates unprecedented challenges in terms of high bandwidth requirements.
This expected growth along with the spectrum crowding encourages the migration to new frequency allocations. In this context, mmWave communications \cite{Rappaport2013millimeter} are emerging as a suitable candidate that can take advantage of a vast amount of unlicensed spectral resource at the mmWave frequency band (over 30-300 GHz) to deal with the high requirements for 5G wireless networks.

With the vision of providing wireless mobile access for UAV-assisted cellular networks in mmWave spectrum, an immediate concern is the extremely high propagation loss, since Friis' transmission law states
that the free space omnidirectional path loss grows with the square of the carrier frequency.
Fortunately, the short wavelength of mmWave signals allows multiple antennas to be  packed into a small UAV  \cite{Pi2011introduction}.
As a byproduct,
beamforming technique can be exploited to construct a narrow directional beam and overcome the high path loss or additional losses caused by atmospheric absorption and scattering. The main difference between a mmWave UAV-assisted cellular network and a conventional mmWave cellular network with a fixed BS is that a UAV-BS may move around. Some of the existing challenges are intensified due to UAV's movement. For example, more efficient beamforming training
and tracking are needed to account for UAV movement, and the channel Doppler effect needs extra consideration, while the UAV position and user discovery are intertwined. This problem has been invoked in recent years.
Table \ref{RmmWavw} shows a summary of the major related works on mmWave UAV-assisted cellular networks.

\begin{table*}
\centering
\setlength{\tabcolsep}{2pt}
\renewcommand{\arraystretch}{1.5}
\extrarowheight 3pt
\caption{Summary of Contributions to mmWave UAV Cellular Networks.}
\begin{tabular}{|l | l | l | l | l | }
\rowcolor{gray!15}
\hline
 \textbf{Reference} & \textbf{Objective} & \textbf{Mobility} & \textbf{Types of BSs} & \textbf{Number of UAVs} \\
\hline
\hline
Xiao \textit{et al.} \cite{xiao2016enabling}  & Beam tracking & Mobile & UAV-only & Single UAV  \\
\hline
Zhu \textit{et al.} \cite{Zhu2018JSAC} & Density optimization  & Static &  UAV-only & Multiple UAVs \\
\hline
Kong \textit{et al.} \cite{Kong2017autonomous} & Optimal relay location & Static & Hybrid (UAVs \& ground BSs) &  Single UAV \\
\hline
Zhao \textit{et al.} \cite{Zhao2018CL}  & Channel tracking & Mobile  & UAV-only &  Single UAV \\
 \hline
Naqvi \textit{et al.} \cite{Naqvi2018Energy}  & Energy efficiency & Mobile  & Hybrid (UAVs, MBS \& SBSs) &  Multiple UAVs \\
\hline
Gapeyenko \textit{et al.} \cite{Gapeyenko2018Effects}  & Optimized deployment & Static  & UAV-only &  Single UAV \\
 \hline
Khosravi \textit{et al.} \cite{Khosravi2018Performance}  & Performance evaluation & Mobile  & UAV-only &  Two UAVs \\
 \hline
Khawaja \textit{et al.} \cite{Khawaja2017UAV}  & Channel measurement & Mobile  & UAV-only &  Single UAV \\
 \hline
\end{tabular}
\label{RmmWavw}
\end{table*}

Benefited by the abundant bandwidth and short wavelength,
Xiao \textit{et al.} \cite{xiao2016enabling} first introduced the concept of mmWave UAV cellular network along with its characteristics and pointed out possible solutions ahead. Specifically, they investigated a hierarchical beamforming codebook structure for fast beamforming training and tracking and explored the use of mmWave spatial-division multiple access for cellular network capacity improvement.
Zhu \textit{et al.} \cite{Zhu2018JSAC}  examined the secrecy performance of a randomly deployed UAV-enabled mmWave communication network over Nakagami-$m$ fading channels, where Mat{\'e}rn Hardcore point process was applied to maintain the minimum safety distance between the randomly deployed UAV-BSs.
In order to bypass these obstacles brought by short wavelength as mentioned above, UAVs as mobile relays are widely needed in mmWave communications. For real applications, it is challenging to find the optimal relay location automatically.
Aiming at this,  Kong  \textit{et al.} \cite{Kong2017autonomous}  studied a novel UAV-relay method specialized for mmWave communications,
where a UAV-relay was used to measure the real-time link qualities and the aerial location was designed properly. The numerical results demonstrated that UAV-relay was capable of providing more accuracy and efficiency solutions than the existing relay method.
An efficient channel tracking method was proposed in \cite{Zhao2018CL} for a mmWave UAV MIMO communication system, where the communication and control system were jointly conceived, and the 3D channel model was formulated as a function of the UAV movement state information and the channel gain information.
In \cite{Naqvi2018Energy}, Naqvi \textit{et al.} formulated a UAV-assisted multi-band HetNet including ground-based macro BS and dual-mode mmWave small BS, and they proposed a joint subcarrier and power allocation scheme to maximize the
system energy efficiency.
By taking into account the LoS blockage by human bodies, Gapeyenko \textit{et al.} \cite{Gapeyenko2018Effects} studied the effective deployment of a mmWave-UAV-BS and derived the corresponding height and cell radius.
In contrast, Khosravi \textit{et al.} \cite{Khosravi2018Performance} evaluated the performance of utilizing small
cell densification technique combined with UAV operating at mmWave frequent band, and the simulation results showed that this method is a promising solution for addressing the propagation limitations.
To top it off, Khawaja \textit{et al.} \cite{Khawaja2017UAV} carried out the propagation measurement for mmWave air-to-ground channels for UAV communications by using ray tracing simulations, four types of environments were analyzed such as urban, suburban, rural, and over sea.

\subsection{UAV NOMA Transmission}

\begin{figure}
\centering
\includegraphics[width=3.0in]{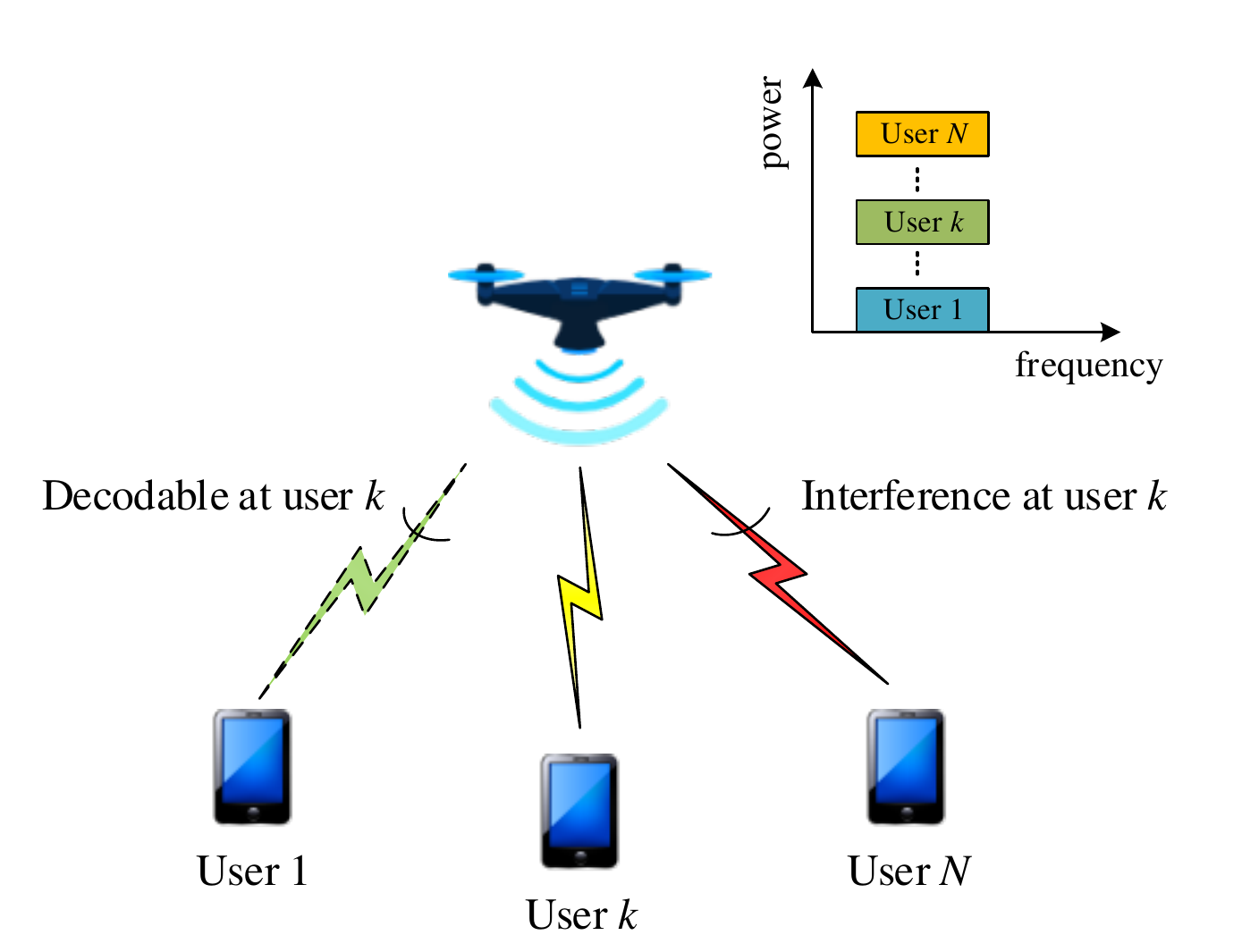}
\caption{A diagram for the considered UAV NOMA scenario. A UAV serves multiple ground users where the signals of users $1$
to $k-1$ are cancelled at the $k$-th user, while the signals of users $k+1$ to
$N$ are received as interference.}
\label{UAV-NOMA}
\end{figure}

NOMA has recently drawn considerable attention as one of the key enabling technologies for 5G communication systems\cite{Ding2017survey},  reaping a high spectral efficiency via incorporating superposition coding at the transmitter with successive interference cancellation (\gls{SIC}) at the receivers.
Compared to orthogonal multiple access schemes (\gls{OMA}), NOMA serves a multitude of users with diversified traffic patterns in a nonorthogonal fashion by considering power domain for multiple access. This provides an effective pathway for UAVs to
ensure the needs of massive ground users at different power levels. The basis of NOMA implementation relies on the difference of channel conditions among users.
Until now, lots of works have contributed to the adoption of NOMA transmission for UAV-assisted communications, in which the UAV-BSs can serve multiple users that operate at the same time/frequency carrier, especially for emergency services with a larger number of users. A simple illustration of the NOMA transmissions in a UAV-based network is depicted in Fig. \ref{UAV-NOMA}.

\begin{table*}
\centering
\setlength{\tabcolsep}{2pt}
\renewcommand{\arraystretch}{1.5}
\extrarowheight 3pt
\caption{Summary of Contributions to UAV NOMA Transmission.}
\begin{tabular}{| l | l | l | l | l | }
\rowcolor{gray!15}
\hline
 \textbf{Reference} & \textbf{Objective} & \textbf{Mobility} & \textbf{Types of BSs} & \textbf{Number of UAVs} \\
\hline
\hline
Sohail \textit{et al.} \cite{Sohail2017maximized}  & Power optimization & Static & UAV-only & Single UAV  \\
\hline
Sohail \textit{et al.} \cite{Sohail2018Access} & Joint power and altitude optimization & Static &  UAV-only & Single UAV \\
\hline
Sharma \textit{et al.} \cite{Sharma2017uav} & Performance analysis & Mobile &  UAV-only &  Single UAV\\
\hline
Baek \textit{et al.} \cite{Baek2017optimal}  & Power optimization & Mobile  & UAV-only &  Single UAV \\

\hline
Rupasinghe \textit{et al.} \cite{Rupasinghe2017non} & Performance analysis & Static & UAV-only & Single UAV   \\

\hline
Rupasinghe \textit{et al.} \cite{Rupasinghe2018non} &  Performance analysis & Static & UAV-only & Single UAV \\
\hline
Rupasinghe \textit{et al.} \cite{Rupasinghe2018comparison} & Performance analysis & Static & UAV-only & Single UAV \\
 \hline
Nasir \textit{et al.} \cite{Nasir2018Enabled} & Max-min rate optimization & Static & UAV-only & Single UAV \\
 \hline
Hou \textit{et al.} \cite{Hou2018Multiple} & Performance analysis & Static & UAV-only & Single UAV \\
 \hline
Liu \textit{et al.} \cite{Liu2018UAVCommunications} & Power allocation and trajectory design & Mobile & UAV-only & Single UAV \\
 \hline
Pan \textit{et al.} \cite{Pan2018JSACPan} & Theoretical-plus-experimental investigation & Mobile & UAV-only & Single UAV \\
 \hline
Pan \textit{et al.} \cite{Nguyen2018JSACNovel} & Radio resource allocation & Mobile & Hybrid (UAVs \& MBS) & Multiple UAV \\
 \hline
\end{tabular}
\label{RNOMA}
\end{table*}

More specifically, for a two-user NOMA case,
the work presented by Sohail and Leow \cite{Sohail2017maximized} focused on
finding the optimal altitude of a rotary-wing UAV-BS to maximize the fairness among users under individual user-rate constraint and the promised gains achievable by NOMA were exhibited over OMA. By extending \cite{Sohail2017maximized}, the authors further formulated the sum-rate problem as a function of power allocation and UAV altitude in \cite{Sohail2018Access} and a constrained coverage expansion methodology was proposed to find the optimal altitude.
Sharma and Kim \cite{Sharma2017uav} adopted a fixed-wing UAV to serve two ground users using downlink NOMA transmission, in which the outage probability for both ground users was derived and an effective transmission mode was selected to guarantee better outage probability for achieving user fairness.
In addition, the NOMA transmission was also applied to the UAV-assisted relaying systems  \cite{Baek2017optimal},
where a novel optimal resource allocation algorithm was proposed to maximize the throughput of ground users and to extend the actual operation range of the UAV.

For a multi-user NOMA case,
Rupasinghe \textit{et al.} \cite{Rupasinghe2017non} introduced the NOMA transmission at UAV-BS operating on mmWave frequency in a large stadium, where multiple users were served simultaneously within the same beam and the optimal operational altitude as well as power
allocation strategy were identified to enhance the outage sum-rate performance.
With the mmWave-NOMA transmission at UAV-BS,  tracking and feeding back full channel state information become
cumbersome, thus the limited feedback schemes were tackled in \cite{Rupasinghe2018non} and \cite{Rupasinghe2018comparison} based on the availabilities of user distance information and user angle information for user ordering. The numerical results revealed that the proposed user angle-based feedback scheme was significantly superior to the proposed user distance-based feedback scheme.
Nasir \textit{et al.} \cite{Nasir2018Enabled} employed a single-antenna UAV-BS and NOMA technique to serve a large number of ground users, where the max-min rate optimization problem was formulated by jointly optimizing multiple parameters (i.e., the UAV's flying altitude, transmit antenna beamwidth, and the amount of power and bandwidth) and a path-following algorithm was developed to solve the non-convex problem.
As a further advance, Hou \textit{et al.} \cite{Hou2018Multiple} proposed an MIMO-NOMA aided UAV framework where a multi-antenna UAV communicates with multiple users equipped with multiple antennas each. By adopting stochastic geometry, the locations of NOMA users were modeled as independent spatial random processes and the closed-form expressions for outage probability of paired NOMA users were derived. With this approach, the positions of UAVs and ground users were modeled in an NOMA-enabled UAV network by Liu \textit{et al.} \cite{Liu2018UAVCommunications} and the system performance was evaluated. Also, they applied a machine learning framework to solve the dynamic placement and
movement of UAVs in a 3D space. Very recently, Pan \textit{et al.} \cite{Pan2018JSACPan} developed a network-coded multiple access downlink scheme for UAV communications, which was more robust against varying downlink channel conditions by the experimental results, while a cooperative NOMA scheme was applied in a wireless backhaul network \cite{Nguyen2018JSACNovel} where UAVs were used as flying small cell BSs to maximize the sum rate of all users, by jointly optimizing the UAVs' positions, the decoding order of the NOMA process and the transmit beamforming vectors.
Finally, Table \ref{RNOMA} portrays a summary of the existing major contributions to UAV NOMA transmission.

As discussed in the above-mentioned literature, it is evident that NOMA is flexible and efficient in multiplexing a number
of end users to UAV communications. However, the successful operation of NOMA in UAV communications requires numerous associated challenges and constraints for the following
reasons:
\begin{itemize}
\item The distinct feature of NOMA with improved spectral efficiencies is that a sophisticated SIC technique at the receiver side is used.
\item SIC exclusively relies on the channel state information at both the receivers and the transmitters to determine the allocated power for each receiver and the decoding order, which needs to be estimated relatively accurately in a UAV communication network.
\item NOMA multiplexing multiple users in the power domain introduces interlayer interference, more efforts are needed to further eliminate the resulting interlayer interference in UAV communications with NOMA.
\item Considering the high mobility of UAVs in practice, the communication distance between the UAV and ground users would vary constantly based on the realtime requirements, thereby the SIC decoding order determined by the received signal strengthes of difference users varies with the locations of UAVs.
\end{itemize}

\subsection{Cognitive UAV Networks}

Nowadays, one crucial predicament faced by the UAV-enabled wireless networks is the shortage of radio spectrum.
Many concerning reasons are listed as follows: i) there is a dramatic growth and usage of new portable mobile devices on the ground (such as smartphones and tablets); ii) different wireless networks (Bluetooth, WiFi, LTE and cellular networks) coexist on the operating spectrum bands of UAVs. These lead to a very intense competition of spectrum usage and thus UAV communications will face the problem of spectrum scarcity \cite{saleem2015integration,Liu2013An}.
Therefore, it is necessary for UAV communications to obtain further spectrum access by dynamic utilization of the existing frequency bands.

Thus far, many researchers and standardization groups have presented the incorporation of CR and UAV communication systems to increase the spectrum opportunities, which is referred to as \textit{cognitive UAV communications} \cite{Ding2018amateur,huang2018cognitive}. This concept constitutes a promising network architecture that allows the coexistence of UAVs with terrestrial mobile devices operating in the same frequency band.
In this case, the UAV-to-ground communications may cause severe interference to the existing terrestrial devices since UAVs usually
have strong LoS links with ground users. Table \ref{RCognitive} shows a number of existing contributions to cognitive UAV networks at a glance.

\begin{table*}
\centering
\setlength{\tabcolsep}{2pt}
\renewcommand{\arraystretch}{1.5}
\extrarowheight 3pt
\caption{Summary of Contributions to Cognitive UAV Networks.}
\begin{tabular}{| l | l | l | l | l | }
\rowcolor{gray!15}
\hline
 \textbf{Reference} & \textbf{Objective} & \textbf{Mobility} & \textbf{Types of BSs} & \textbf{Number of UAVs} \\
\hline
\hline
Huang \textit{et al. } \cite{huang2018cognitive} & Joint trajectory and power optimization & Mobile & Hybrid (UAV \& ground BSs) & Single UAV  \\
\hline
Zhang \textit{et al. } \cite{zhang2017spectrum}& Optimal deployment & Static &  Hybrid (UAVs \& ground BSs) & Multiple UAVs \\
\hline
Sboui \textit{et al. } \cite{Sboui2017energy} & Power optimization & Static &  Hybrid (UAVs \& ground BS)  &  Two UAVs\\
 \hline
\end{tabular}
\label{RCognitive}
\end{table*}

Note that in the literature, there have been several works that studied the cognitive UAV communication system.
For example, \cite{huang2018cognitive} jointly optimized the UAV's trajectory and transmit power allocation with the aim of achieving the maximum throughput of a cognitive UAV communication, while restricting the interference imposed at primary receivers below a tolerable level.
Zhang and Zhang \cite{zhang2017spectrum}  presented an underlay spectrum sharing  method between the
drone-cells network and traditional ground cellular network under different scenarios, i.e.,  spectrum sharing of single tier drone-cells in 3D network, and a spectrum sharing between the drone-cells network and the traditional two-dimensional (\gls{2D}) cellular network.
Using stochastic geometry theory, they derived the explicit expressions for the drone-cells coverage probability and achieved the optimal density of UAV-BSs for maximizing the throughput.
Similarly,  Sboui \textit{et al.} \cite{Sboui2017energy} proposed to integrate an underlay CR into a UAV system where the UAV as a secondary transmitter opportunistically exploited and shared the primary spectrum for the UAV-to-ground transmission. The objective was to maximize the energy efficiency of the UAV unit, thereby ensuring effective and long-time operations of UAVs.

\subsection{Energy Harvesting UAV Networks}

Unlike traditional ground transceivers connected to external power supplies, UAV is powered by capacity-limited battery and thus the UAV-based communications are facing the limited energy availability for performing various operations like flight control, sensing/transmission of data or running some applications.
As is known to all, the finite on-board energy storage of typical UAVs (battery life is usually less than 30 minutes) restricts their operation time (i.e., flight time or hovering time) \cite{Long2018energy}, and it is not always possible that the UAVs are required to return to the depot for
battery charging frequently. Thus, this is critical but challenging to guarantee stable and sustainable communication services and will act as a performance bottleneck.

\subsubsection{Energy Efficiency}
For many UAV applications, energy consumption saving is of significant importance to prolong the lifetime of a UAV network.
In recent years, many research endeavors have been conducted on the energy-aware UAV deployment and operation mechanisms.  More explicitly,
Li \textit{et al.} \cite{Li2016TMCEE} proposed an energy-efficient transmission scheduling scheme of UAVs in a cooperative relaying network such that the maximum energy consumption of all the UAVs was minimized, in which
an applicable suboptimal solution was developed and the energy could be saved up to $50\%$ via simulations.
By exploiting the optimal transport theory, Mozaffari \textit{et al.} \cite{Mozaffari2016ICCOptimal} investigated the energy-efficient deployment of multiple UAV-BSs
for minimizing the total required transmit power of UAVs under the rate requirements of the ground users.
In \cite{Zeng2018EETWC}, Zeng \textit{et al.} presented an energy-efficient UAV communication
by optimizing UAV's trajectory with a fixed altitude, where the propulsion energy consumption of the fixed-wing UAV was taken into account and the theoretical model was derived.
Ghazzai \textit{et al.} \cite{Ghazzai2017TGCN} developed an energy-efficient optimization problem for a UAV communication by integrating
CR technology to minimize the total energy consumption of UAV including the flying and communication energies, where
a joint algorithm inspired from the Weber formulation was proposed to optimize the transmit power level and the location of
cognitive UAV. Liu \textit{et al.} \cite{Liu2018UAVJSAC} proposed a framework that leveraged deep reinforcement learning to study the energy consumption
used for UAV movements, while maintaining the fair communication coverage and the network connectivity.
Ruan \textit{et al.} \cite{Ruan2018China} built a multi-UAV energy-efficient coverage deployment model, in which the proposed model was
decomposed into two subproblems to reduce the complexity of strategy selection, i.e., coverage maximization and power minimization.

\begin{figure}
\centering
\includegraphics[width=2.5in]{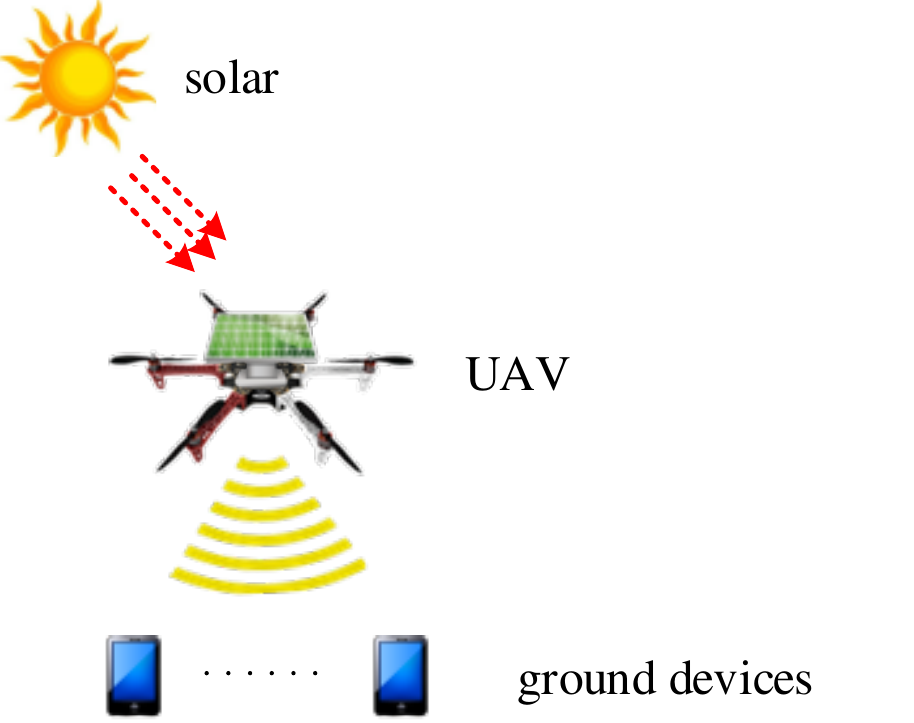}
\caption{A solar-powered UAV communication system, where the UAV is equipped with solar panels that can harvest energy from solar source.}
\label{solar-powered}
\end{figure}

\subsubsection{Energy Harvesting}

In fact, the energy consumption of the battery-powered UAV is usually split into an energy consumed by the communication unit and the energy used for the hardware and mobility of UAVs. Hence, energy harvesting UAV is crucial to  prolong its flight duration without adding significant
mass or size of the fuel system. In recent applications, it is very advantageous to harvest energy from ambient sources for recharging UAV's
battery, which is referred to as \textit{wireless powered UAV networks}. A lot of related works have proceeded to improve the endurance of electrically powered UAVs. In particular, solar-powered UAV has received significant attention that harvests energy from solar and converts it to electrical energy via photovoltaic effect for realizing perpetual flight, the system model is shown in Fig. \ref{solar-powered}.
As a matter of fact, the available solar energy depends on the geographic location, altitude, the number of daylight hours and the day of the year. The solar-powered UAV prototypes have been developed by engineers in \cite{Morton2015} and \cite{Oettershagen2016} and they
revealed the possibility of continuous flight for $28$ hours. In \cite{Wang2018AccessAnalysis}, Wang \textit{et al.} constructed the simulation model of solar cell for solar-powered UAV via MATLAB/Simulink software and  the output characteristics of solar panels in three types of weather conditions (i.e., light rain, cloudy and sunny days) were tested in Nanjing, the test site was Nanjing University of Aeronautics and Astronautics. The experimental test results demonstrated that the output curve of a solar cell was mainly affected by the ambient temperature and light intensity.
Indian Institute of Technology Kanpur \cite{Dwivedi2018India} carried out a day-only flight test of a solar UAV platform in April 2017, where the UAV
took off at 9:30 am successfully landed at 6:00 pm. From the measurement results, one could observe that the generated power became less than the power required by the UAV system during around from 3:30 pm to 5:10 pm. Meanwhile, the Aircraft Design Group of Cranfield University at United
Kingdom \cite{Rajendran2018Rajendran } examined the impact of temperature and solar irradiance intensity on various solar module angles, the results showed that the optimum operating temperature for both non-laminated and laminated solar modules was around $45^\circ$C and the solar power rose almost linearly along the solar module tilt angle.

\begin{table*}
\centering
\setlength{\tabcolsep}{2pt}
\renewcommand{\arraystretch}{1.5}
\extrarowheight 3pt
\caption{Summary of Contributions to Energy Harvesting UAV Networks.}
\begin{tabular}{| l | p{4.5cm} | l | l | l | }
\rowcolor{gray!15}
\hline
 \textbf{Reference} & \textbf{Objective} & \textbf{Mobility} & \textbf{Types of BSs} & \textbf{Number of UAVs} \\
\hline
\hline
Sun \textit{et al. } \cite{Sun2018resource} & Joint optimizing position and power and subcarrier & Static & UAV-only & Single UAV  \\
\hline
Sun \textit{et al.} \cite{Sun2018OarXivOptimal} & Joint optimizing 3D-trajectory and power and subcarrier & Mobile & UAV-only & Single UAV  \\

\hline
Hua \textit{et al. } \cite{Hua2017throughput} & Optimal location & Mobile &  Hybrid (UAV \& BS) & Single UAV \\
\hline
Wu \textit{et al. } \cite{Wu2018pathplanning} & Path planning  & Mobile &  UAV & Single UAV \\
\hline
Sowah \textit{et al. } \cite{Sowah2017rotational} & Implementation viewpoint& Static &  UAV-only &  Single UAV\\
\hline
Long \textit{et al. } \cite{Long2018energy}  & Routing protocol & Mobile  & Hybrid (UAVs \& ground BSs) &  Multi-tier UAVs \\
\hline
Yang \textit{et al.}  \cite{Yang2018CLOutage} & Outage Performance & Static  & Hybrid (UAV \& ground BS) &  Single UAV \\
 \hline
Marano \textit{et al. } \cite{Marano2018resource}  & Energy allocation & Mobile  & UAV-only &  Single UAV \\
 \hline
Wang \textit{et al. } \cite{Wang2018resource}  & Power optimization and time allocation & Static   & UAV-only &  Single UAV \\
 \hline
Xu \textit{et al. } \cite{Xu2017uav} & Trajectory optimization & Mobile   & UAV-only &  Single UAV \\
 \hline
Xu \textit{et al. } \cite{Xu2018UAV}  &  Trajectory optimization & Mobile   & UAV-only &  Single UAV \\
 \hline
Nguyen \textit{et al. } \cite{Nguyen2018WCL}  & Optimal energy harvesting time and power control & Static  & UAV-only &  Single UAV \\
 \hline
Park \textit{et al. } \cite{Park2018minimum}  & Joint trajectory optimization and resource allocation & Mobile  & UAV-only &  Single UAV; Two UAVs \\
 \hline
Xie \textit{et al. } \cite{Xie2018throughput}  & Joint trajectory optimization and resource allocation& Mobile  & UAV-only &  Single UAV \\

 \hline
\end{tabular}
\label{REnergy}
\end{table*}

From the standpoint of academic research, Sun \textit{et al.} \cite{Sun2018resource} have invoked the resource allocation
design for a solar-powered multicarrier UAV communication system for maximization of the system sum throughput, where a low-complexity
joint 3D position, power and subcarrier allocation algorithm was proposed to find out the suboptimal solution.
Since the aerodynamic power consumption of realistic UAV systems depends on the flight velocity, the assumption of constant
aerodynamic power consumption is not valid in practice. For this reason, Sun \textit{et al.} \cite{Sun2018OarXivOptimal} further
studied a multicarrier solar-powered UAV communication system by jointly taking into account the solar energy harvesting,
the aerodynamic power consumption, the dynamics of the on-board energy storage, and the QoS requirements of the ground users. The objective was to maximize the system sum throughput over a given time period.
Simulation results showed that the UAV could harvest more solar energy when it was flying right above the clouds.
Hua \textit{et al.} \cite{Hua2017throughput} considered an energy-constrained UAV relaying scenario where the power splitting-based relaying
protocol was adopted at a UAV for energy harvesting and information processing with the aim of maximizing the network
throughput. In urban environment, Wu \textit{et al.} \cite{Wu2018pathplanning} proposed a solar-powered UAV path planning framework that considered the obstacle condition and the shadow regions caused by high buildings.
However, the solar energy for solar cell-based harvesting is often weather-dependent and unpredictable, thereby suffering from uncertainty caused by random energy arrivals. Most current works did not consider this practical environment.
In this context, the work by Sowah \textit{et al.} in \cite{Sowah2017rotational} presented a rotational energy harvester based on a brushless dc generator to harvest ambient energy for prolonging the indoor flight time of quadcopter, while a prototype of the rotational energy harvesting system was also implemented.
Long \textit{et al.}  \cite{Long2018energy} proposed the architecture of energy neutral internet of UAVs where recharging stations were used to energize the UAVs via wireless power transfer (\gls{WPT}) with radio frequency (\gls{RF}) signals, which significantly enabled the continuous operation lifetime.
In UAV-assisted relaying systems, Yang \textit{et al.}  \cite{Yang2018CLOutage} analyzed the outage performance of UAV harvesting
energy from the ground BS, where both the shadowed-Rician fading and shadowed-Rayleigh fading were respectively considered.

\begin{figure}
\centering
\includegraphics[width=3.3in]{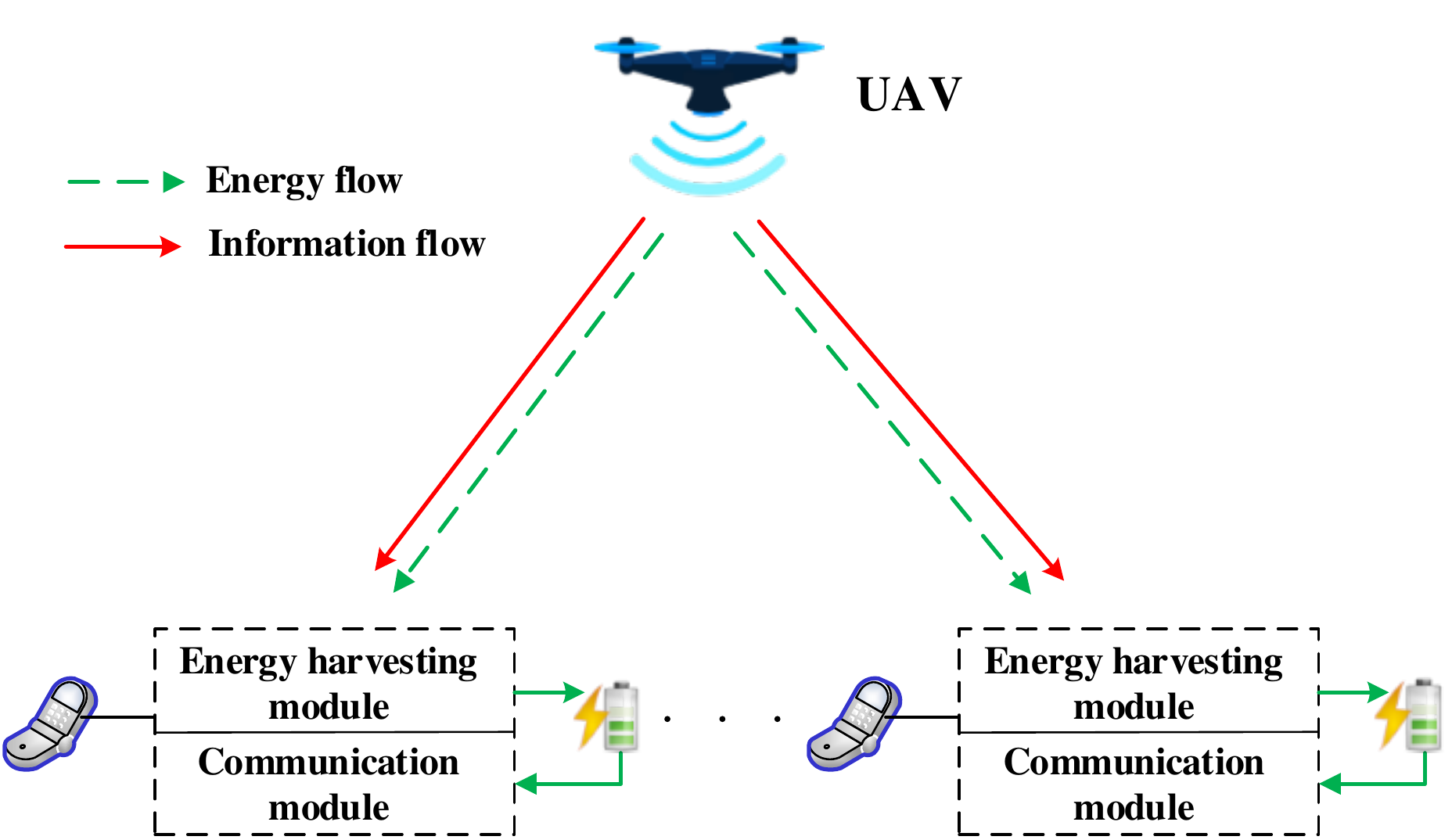}
\caption{A typical example of a UAV-enabled wireless powered network, where the UAV is able to transmit energy or simultaneously transmit data and energy to ground user devices via RF signals. The green color portion represents the energy flow and the red color portion represents the information flow.}
\label{UAV-EH}
\end{figure}

On the other hand, mobile devices (such as low-lower sensors) usually are also energy-constrained and the useful lifetimes are limited by the battery capacity. Since UAVs have more energy available than mobile devices and the UAVs actually provide services to the ground devices, UAVs as aerial energy transmitters with additional flexibility are expected  to provide ubiquitous wireless energy supply to massive low-power devices. This significantly improves the wireless charging efficiency compared to the conventional ground charging stations at fixed locations,
which is referred to as \textit{UAV-enabled wireless powered networks}.
A detailed example of the energy harvesting from UAV is presented in Fig. \ref{UAV-EH}, where the UAV comprises an energy
transfer module for broadcasting RF energy to ground user devices.
The idea has been conducted in recent years \cite{Marano2018resource,Wang2018resource,Xu2017uav,Xu2018UAV,Nguyen2018WCL}. To be specific,
Marano and Willett \cite{Marano2018resource} looked into improving the sustainable operation of sensors during the sensing stage in the context of a wireless sensor network via RF signal from a UAV. Wang \textit{ et al.} \cite{Wang2018resource} proposed a joint time and power optimization algorithm for maximization of the average throughput, where UAV acted as a static energy source to power multiple D2D pairs and a harvest-transmit-store protocol was adopted.  However, this work did not take into account the mobility of UAV. By exploiting UAV's trajectory design, Xu \textit{et al.}
\cite{Xu2017uav} presented the first work on characterizing the achievable energy region of ground users in a UAV-enabled two-user WPT system.
The authors of \cite{Xu2017uav} further considered the UAV-enabled multiuser WPT system in \cite{Xu2018UAV}, where the problems of sum-energy maximization and min-energy maximization were respectively conceived by optimizing the UAV's trajectory subject to the practical speed constraint. Subsequently, Nguyen \textit{et al.} \cite{Nguyen2018WCL} considered the energy efficiency problem in WPT-powered D2D communications with the help of UAV by jointly optimizing the energy harvesting time and power allocation, and then the performance of the UAV
network was evaluated by embedded optimization module implemented in Python.

As a further development, Park \textit{et al.}
\cite{Park2018minimum} have invoked a UAV-aided wireless powered communication network (\gls{WPCN}) for maximization of the minimum user throughput by jointly optimizing the UAV's trajectory, uplink power control and time resource allocation, where both the scenarios of integrated UAV and separated UAV WPCNs were respectively taken into consideration.
In \cite{Xie2018throughput}, Xie \textit{et al.}  addressed a joint UAV trajectory and resource allocation optimization problem for the uplink throughput maximization in a UAV-enabled WPCN setup, while maintaining the UAV's maximum speed constraint and the users' energy neutrality constraints.
Eventually, a summary of the existing works on energy harvesting UAV networks is shown in Table \ref{REnergy}.

\section{Network Layer Techniques \label{d}}
The next generation networks should intelligently and seamlessly integrate multiple nodes to form a multi-tier hierarchical architecture, including the drone-cell tiers for large radio coverage areas, the ground small cell tiers for small radio coverage areas, the user device tiers with D2D communications, and so forth. However, the integration of different tiers will result in new issues to the investigation of the network layer techniques.  Therefore, specific strategies that coordinate the QoS of nodes are necessary.
In this section, we review the state-of-the-art works on UAV-assisted HetNets, combined UAVs and D2D communications, and software defined UAV  networks.

\subsection{UAV-Assisted HetNets}

With the forthcoming of 5G era, densely populated users are thirsty for broadband wireless communications and network
operators are expected to support diverse services with high wireless data demands such as multimedia streaming and
video downloads. The unrelenting increment in mobile traffic volumes imposes an unacceptable burden on the operators in
terms of increased capital expenditure and operating costs. An intuitive option to offload the cellular traffic is to deploy small
cells (e.g., pico and femto cells). However, in unexpected or temporary events, the deployment of terrestrial infrastructures is challenging
since the mobile environments are sophisticated, volatile, and heterogeneous. One potential solution resorts to the usability
of drone-cells  \cite{Mozaffari2018tutorial}, which has been proved to be instrumental in supporting ground cellular networks in areas of
erratic demand. The idea is to bring the ground users closer to the drone-cells in order to improve their QoS due to the short-range LoS connections from sky.
Fig. \ref{UAVasBSs} shows the typical UAV-assisted HetNet architecture with one MBS and multiple drone-cells.

\begin{table*}
\centering
\setlength{\tabcolsep}{2pt}
\renewcommand{\arraystretch}{1.5}
\extrarowheight 3pt
\caption{Summary of Contributions to UAV-Assisted HetNets.}
\begin{tabular}{| l | l | l | l | l | }
\rowcolor{gray!15}
\hline
 \textbf{Reference} & \textbf{Objective} & \textbf{Mobility} & \textbf{Types of BSs} & \textbf{Number of UAVs} \\
\hline
\hline
Li \textit{et al.} \cite{Li2017uav} & Optimized bandwidth & Static & Hybrid (UAVs \& MBS) & Multiple UAVs  \\
\hline
Kumbhar \textit{et al.} \cite{Kumbhar2018exploiting} & Optimal placement & Static &  Hybrid (UAVs \& MBSs) & Multiple UAVs \\
\hline
Merwaday \textit{et al.} \cite{Merwaday2016improved} & Optimal location & Mobile &  Hybrid (UAVs \& MBSs) &
Multiple UAVs\\
\hline
Sharma \textit{et al.} \cite{Sharma2016uav}  & Optimal placement & Static  & Hybrid (UAVs \& MBS) &  Multiple UAVs \\
 \hline

Zhang \textit{et al.} \cite{Zhang2018machine}  & Optimal placement & Mobile & Hybrid (UAVs \& ground BSs)  &  Multiple UAVs \\
 \hline
Sharma \textit{et al.} \cite{Sharma2016uavs}  & Optimal placement & Static & Hybrid (UAVs \& MBSs) &  Multiple UAVs \\
 \hline
Sun \textit{et al.} \cite{Sun2017Latency} & Optimal placement & Static  & Hybrid (UAV \& MBS) &  Single UAV \\
 \hline
Mehta \textit{et al.} \cite{Mehta2017aerial}  & Optimal placement & Static  & Hybrid (UAVs \& MBS) &  Multiple UAVs \\
 \hline
Sharma \textit{et al.} \cite{Sharma2017intelligent}  & Optimal placement & Static  & Hybrid (UAVs \& MBS) &  Multiple UAVs \\
 \hline
Sekander \textit{et al.} \cite{Sekander2018multi}  & Altitude optimization&Static & Hybrid (UAV \& BSs) &  Two-tier UAVs \\
 \hline
\end{tabular}
\label{RHetNets}
\end{table*}

At the same time, the mobility of drone-cells enables them to serve users with high mobility and data rate demand.
In the open literature, two canonical lines of research can be identified involving \textit{ground HetNets} and \textit{aerial HetNets}.  In the first line focusing on characterizing the ground HetNets,
Li and Cai \cite{Li2017uav}  introduced UAV-based floating relays to handle the increasing traffic volume due to the rapid
development of mobile Internet, where UAVs were parked inside a small garage on the MBS and their batteries would be recharged when backing to the garage. The optimized bandwidth solution was proposed to enable heterogeneous deployment of UAV-based floating
relay cells inside the macro cell and achieved dynamic and adaptive coverage.
In a two-tier UAV-assisted HetNet, Kumbhar \textit{et al.} \cite{Kumbhar2018exploiting} conceived the optimal deployment of UAVs and the interference coordination technique defined in LTE-Advanced was exploited to mitigate the inter-cell interference resulting from the HetNet.
At the same time, the genetic algorithm was shown to be an effective method to maximize the spectral efficiency of the network.
Merwaday \textit{et al.} \cite{Merwaday2016improved} aimed at exploring a large-scale disaster-affected environment consisting of MBSs and small cell base station (\glspl{SBS}) in which UAVs were designated for providing coverage and seamless broadband connectivity in desired regions.
A genetic algorithm was proposed to optimize the positions of UAV-BSs with the goal of maximizing the network throughput.

As a further advance, the problem of user-demand-based UAV assignment in HetNets was investigated in \cite{Sharma2016uav},
a neural-based cost function framework was formulated to strike the appropriate user demand areas and UAVs to enhance the network capacity.
Zhang \textit{et al.} \cite{Zhang2018machine} considered a heterogeneous cellular network comprising a set of UAVs as flying BSs and a set of ground BSs that provided an on-demand wireless service to a group of cellular users. They developed  a novel machine learning framework to predict the cellular data traffic and formulated a power minimization problem for downlink communications and mobility to optimize the deployment of UAVs.
Furthermore, the minimum delay scheme has been widely studied in UAV-assisted HetNets for improving QoS of mobile users.
For example, in \cite{Sharma2016uavs} the concept of entropy nets from the neural network was applied to minimize the overall network delay by optimizing the placement and distribution of cooperating UAVs in demand areas.
Sun and Ansari \cite{Sun2017Latency} tried to balance the traffic loads between a UAV-BS and an MBS for achieving a minimum total average latency ratio among the MUs under the energy limitations of the UAV-BS. The optimization problem formulated comprised two steps, namely
first determining the location of the UAV-BS and then optimizing the association coverage of the UAV-BS.

Another line of research has established that future aerial networks will be heterogeneous and comprise different types of UAVs, namely high-altitude long-range UAVs  (less than 5km), medium-altitude UAVs  (between 5km to 10km), low-altitude short-range UAVs  (greater than 10km) \cite{Shah2017distributed}.
The multi-tier aerial networks are much affected by the density of users and services and can be constructed by utilizing several UAV types, which is similar to terrestrial HetNets with macro-, small-, pico-cells, and relays.
As an initial study, Mehta and Prasad \cite{Mehta2017aerial} introduced the concept of aerial-HetNet to offload the data traffic from the congested ground BSs in hotspots, where a fleet of small UAVs were deployed as an ad-hoc network with variable operational altitudes in the air. The network performance improvement was also illustrated.
To cater for the capacity and coverage enhancements of HetNets, an MBS-based decisive and cooperative problem was presented in \cite{Sharma2017intelligent} for the accurate mapping of the UAVs to the demand areas, where both the single-layer model with multiple
UAVs and the multi-layer model with multiple UAVs in each layer were respectively considered.  An intelligent solution utilizing the priority-wise dominance and the entropy approaches was proposed for the accurate and efficient placement of the UAVs.
Sekander \textit{et al.} \cite{Sekander2018multi} concentrated on investigating the feasibility of multi-tier UAV network architecture over traditional single-tier UAV network in terms of spectral efficiency of downlink transmission, and identified  the relevant challenges such as energy consumption
of drones, interference management and so forth.  The impact of different urban environments (including high-rise urban, suburban,
and dense urban) on this multi-tier UAV network architecture was finally shown by numerical results.
Finally, a brief summary of the above contributions is given in Table \ref{RHetNets}.

\subsection{Combined UAVs and D2D Communications}

\begin{figure}
\centering
\includegraphics[width=3.0in]{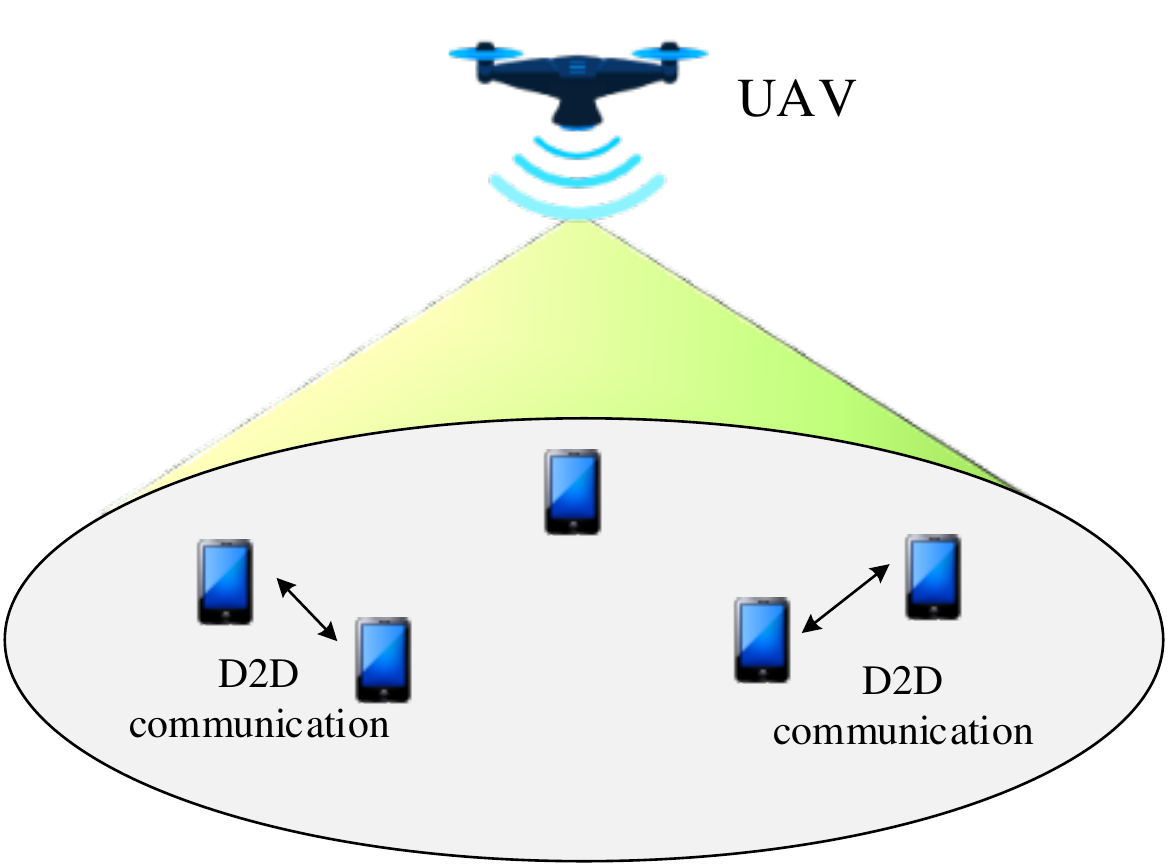}
\caption{D2D communications underlying UAV-supported cellular network.}
\label{UAV-D2D}
\end{figure}

D2D communications as a new network architecture is becoming increasingly popular,
which dramatically improves network capacity by offloading mobile traffic from BSs, when two neighboring nodes communicate with each other via
D2D mode. In general,
D2D communications are typically deployed using underlay transmission links which reuse existing licensed spectrum resources \cite{Jameel2018survey}, while UAV can be a good candidate to promptly construct the D2D-enabled wireless network by introducing a new dimension, as shown in Fig. \ref{UAV-D2D}. In parallel, the working of UAVs alongside D2D communications over a shared spectrum band will also introduce important interference management challenges,  thus the impact of UAV's mobility on D2D and network performance should be analyzed.
Table \ref{RD2D} shows a summary of the existing major contributions to combined UAVs and D2D communications.

\begin{table*}
\centering
\setlength{\tabcolsep}{2pt}
\renewcommand{\arraystretch}{1.5}
\extrarowheight 3pt
\caption{Summary of Contributions to Combined UAVs and D2D Communications.}
\begin{tabular}{| l | l | l | l | l | }
\rowcolor{gray!15}
\hline
 \textbf{Reference} & \textbf{Objective} & \textbf{Mobility} & \textbf{Types of BSs} & \textbf{Number of UAVs} \\
\hline
\hline
Mozaffari \textit{et al.} \cite{mozaffari2016unmanned} & Performance analysis & Static;  Mobile & UAV-only & Single UAV  \\
\hline
Tang \textit{et al.} \cite{Tang2018AC} & Channel Assignment & Mobile &  UAV-only & Multiple UAVs \\
\hline
Guo  \textit{et al.} \cite{guo2017coverage} &  Altitude optimization & Static &  UAV-only &  Multiple UAVs \\
\hline
Christy \textit{et al.} \cite{christy2017optimum}  & Trajectory optimization & Mobile & UAV-only &  Single UAV \\
 \hline
Wang \textit{et al.} \cite{Wang2018FDUAV}  & Spectrum sharing planning & Mobile & UAV-only &  Single UAV \\
 \hline
Xue \textit{et al.} \cite{Xue2018Access}  &  Social group utility maximization & Static & UAV-only &  Single UAV \\
 \hline
\end{tabular}
\label{RD2D}
\end{table*}

To elaborate, Mozaffari \textit{et al.} \cite{mozaffari2016unmanned} conducted the first attempt on providing a comprehensive performance analysis to evaluate the coexistence of UAV and D2D in terms of different performance metrics, in which  both key scenarios namely static UAV and mobile UAV were considered, respectively.
Tang \textit{et al.} \cite{Tang2018AC} studied the assignment of the radio channels in a combined UAV and D2D-based network with the consideration of high mobility of UAV and D2D nodes, in which the UAVs could be used as both local content servers and aerial D2D nodes. Moreover, a distributed anti-coordination game algorithm was conceived for solving the channel assignment problem.
In a multi-UAVs-enabled wireless network with D2D communications, Guo \textit{et al.} \cite{guo2017coverage} provided an analysis of the coverage probability of downlink users and D2D users and then optimized the altitude of UAVs to maximize the capacity of ground network.
Christy \textit{et al.} \cite{christy2017optimum} examined the utilization of a UAV to discover potential D2D devices for establishing D2D transmissions as an emergency communication network. Through simulation results, the authors have shown that it can reduce the device energy consumption and increase the capacity of the network. The concept of full-duplex was introduced by Wang \textit{et al.} \cite{Wang2018FDUAV} to UAV-assisted
relaying systems with underlaid D2D communications, in which the
transmit power and UAV's trajectory were jointly designed to achieve
efficient spectrum sharing between aerial UAV and terrestrial D2D communications.
Beyond that, the work
in \cite{Xue2018Access} proposed to apply multiple D2D peers to the
UAV-supported social networking for the maximization of the sum social group utility, where both physical interference and social connections between users in the physical/social domain were considered.

\subsection{Software Defined UAV  Networks}

Recent proposals for future wireless network architectures aim to create a flexible network with
improved agility and resilience. SDN has been introduced in 2008 to program the network via a logically software-defined controller \cite{Bera2017software}, which can decouple the control plane and data plane to facilitate network reconfiguration. This is conducive to manage the infrastructure and resources of wireless networks.
Compared to traditional networking, SDN has better controllability and visibility for network components, which enables better management by
using the common controller.

\begin{table*}
\centering
\setlength{\tabcolsep}{2pt}
\renewcommand{\arraystretch}{1.5}
\extrarowheight 3pt
\caption{Summary of Contributions to Software Defined UAV  Networks.}
\begin{tabular}{| l | l | l | l | l | }
\rowcolor{gray!15}
\hline
 \textbf{Reference} & \textbf{Objective} & \textbf{Mobility} & \textbf{Types of BSs} & \textbf{Number of UAVs} \\
\hline
\hline
Bor-Yaliniz \textit{et al.} \cite{Bor2016new} & Optimal placement & Static & Hybrid (UAVs \& ground BSs) & Multi-tier UAVs \\
\hline
Sharma \textit{et al.} \cite{Sharma2017efficient} & Fast handovers & Mobile &  Hybrid (UAVs \& ground BSs) & Multiple UAVs \\
\hline
Rahman \textit{et al.} \cite{Ur2017deployment} & Optimal placement & Static &  UAV-only &  Multiple UAVs \\
\hline
Shukla \textit{et al.} \cite{Shukla2018software}  &  Computation offloading& Static  & Hybrid (UAVs \& ground BSs) &  Multiple UAVs \\
 \hline
Yang \textit{et al.} \cite{Yang2017proactive}  & Optimal density and location & Static  & UAV-only &  Multiple UAVs \\
 \hline
Secinti \textit{et al.} \cite{Secinti2017resilient}  & Multi-path routing & Static & UAV-only &  Multi-tier UAVs \\
 \hline
Secinti \textit{et al.}  \cite{Secinti2018sdns} & Multi-path routing & Static  & UAV-only &  Multiple UAVs \\
 \hline
\end{tabular}
\label{RSDN}
\end{table*}

In real-world applications of drone-cells, wireless networks must be configured efficiently for seamless integration/disintegration of  UAVs, such as changing protocols and creating new paths.
Based on the SDN architecture, UAVs can perform as SDN switches on data plane for collecting context information in a distributed way, while the ground BSs are controllers gathering data and making control decisions on network functions and resource allocation.
Helped by SDN, network reconfiguration and resource allocation among a swarm of UAVs can
be conducted in a more flexible way. Table \ref{RSDN} shows a summary of the existing major contributions to SDN with UAVs.

Pioneering work by Bor-Yaliniz and Yanikomeroglu \cite{Bor2016new} proposed a drone-cell management framework enabled by SDN and network functions virtualization technologies to assist a terrestrial HetNet.
For an SDN-based UAV architecture, the proposed SDN variant in \cite{Sharma2017efficient}  provided handover facilities in the UAVs  (i.e., as on-demand forwarding switches) that supported wireless networks with lower handover latency.
Since SDN can enable a global view of network,  Rahman \textit{et al.} \cite{Ur2017deployment} considered the
placement of the SDN controller in an SDN-based UAV network for providing better service, where  an appearing trade-off was achieved between
the communication overhead and the end-to-end delay for sharing the control information between UAVs and SDN controller.
Shukla \textit{et al.} \cite{Shukla2018software} studied the resource allocation of a multi-UAV network
to minimize the operating delay and energy consumption by considering the edge servers and cloud servers. The network management between these units was enabled by SDN controller in an efficient manner such that the QoS demands of applications were ensured.
Yang \textit{et al.} \cite{Yang2017proactive} developed a proactive UAV-cell deployment framework to alleviate overload
conditions caused by flash crowd traffic. Under this frame, the SDN technology was employed to seamlessly integrate and disintegrate
drone-cells by reconfiguring the network. Similarly, Secinti \textit{et al.} \cite{Secinti2017resilient} carried out  a study on a resilient multi-path routing framework for a UAV-network, where the SDN controller was utilized to determine the preferred routes subjected to jamming.
Also, Secinti \textit{et al.} \cite{Secinti2018sdns} further proposed
an aerial network management protocol building on top of an SDN architecture, where each UAV became a software
switch that performed control directives sent by a centralized controller. Finally, the multi-path routing algorithm was proposed to reduce the average end-to-end outage rate.

\section{Joint Communication, Computing, and Caching \label{e}}

\begin{figure}
\centering
\includegraphics[width=2.3in]{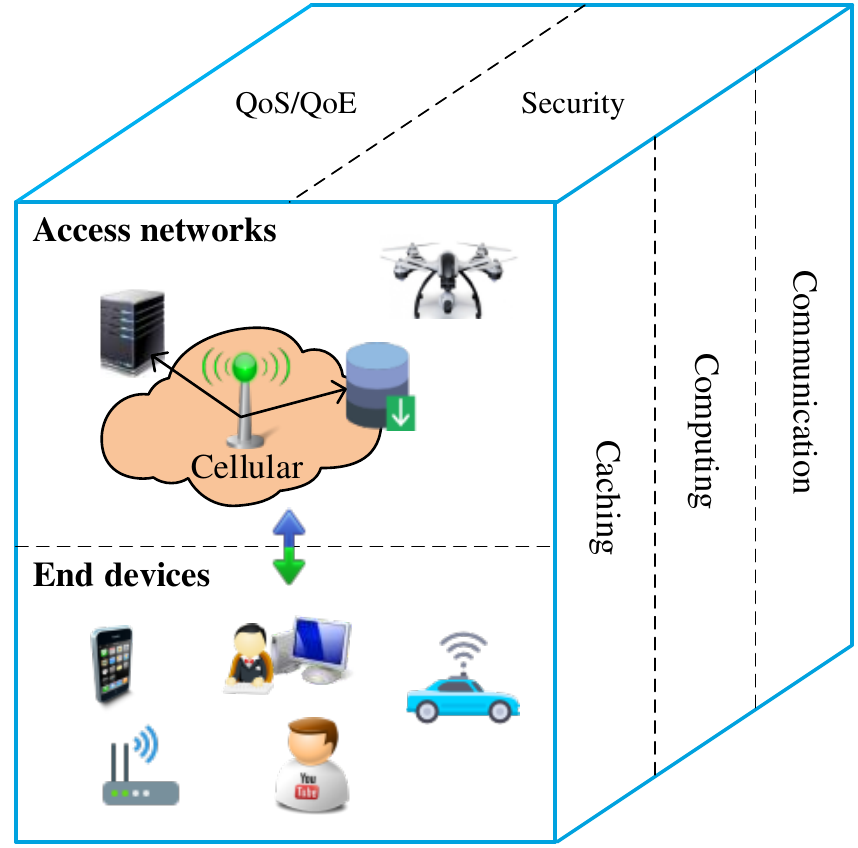}
\caption{Architecture of UAV-oriented communication, computing and caching.}
\label{Joint}
\end{figure}

The 5G wireless network is envisaged to embed various resources to support massive traffic and various services. This will be characterized by the
convergence of communications, computing, and caching capabilities \cite{Wang2017integration}.
As an essential component of IoT and future 5G networks, UAV can not only act as an edge computing platform for providing flexible and resilient
services to IoT devices with limited processing capabilities, but also act as a complementary method to cache some popular contents for reducing backhaul workload and transmission latency at peak time. The architecture of UAV-oriented communication, computing and caching is shown in Fig. \ref{Joint}.
In this section, we review recent works on UAV-based MEC and UAV-based cache, which may be applied in 5G/B5G communications.

\subsection{UAV-Based MEC}

Due to the limited battery and low computation capability, it is challenging for IoT devices
to execute real-time applications. Fortunately, MEC has recently emerged as a paradigm to tackle this issue \cite{Abbas2017mobile}. With the deployment of MEC server,
mobile users can offload their computation tasks to the edge of network by empowering the cloud computing functionalities.
It serves two important purposes:
\begin{itemize}
\item Reduction in application latency (i.e., execution time), if a remote device has enormous computing resources.
\item Improving battery performance because application is being executed at a remote device.
\end{itemize}

\begin{table*}
\centering
\setlength{\tabcolsep}{2pt}
\renewcommand{\arraystretch}{1.5}
\extrarowheight 3pt
\caption{Summary of Contributions to UAV-based MEC.}
\begin{tabular}{| l | l | l | l | l | }
\rowcolor{gray!15}
\hline
 \textbf{Reference} & \textbf{Objective} & \textbf{Mobility} & \textbf{Types of BSs} & \textbf{Number of UAVs} \\
\hline
\hline
Jeong \textit{et al.} \cite{Jeong2017IET} & Computation offloading & Mobile & UAV-only & Single UAV \\
\hline
Jeong \textit{et al.} \cite{Jeong2017mobile} & Computation offloading & Mobile &  UAV-only & Single UAV \\
\hline
Tang \textit{et al.} \cite{Tang2018novel} & Computation offloading & Mobile &  UAV-only &  Multiple UAVs \\
\hline
Zhou \textit{et al.} \cite{Zhou2018uav}  & Computation offloading & Mobile  & UAV-only &  Single UAV \\
 \hline
Jung \textit{et al.} \cite{Jung2017acods}  & Computation offloading & Mobile  & UAV-only &  Single UAV \\
 \hline
Motlagh \textit{et al.} \cite{Motlagh2017uav}  & Computation offloading & Static  & UAV-only &  Single UAV \\
 \hline
 Hua \textit{et al.} \cite{Hua2018EnergyO}  & Computation offloading & Mobile  & Hybrid (UAVs \& ground BSs) &  Multiple UAVs \\
 \hline
\end{tabular}
\label{RMEC}
\end{table*}

\begin{figure}
\centering
\includegraphics[width=2.8in]{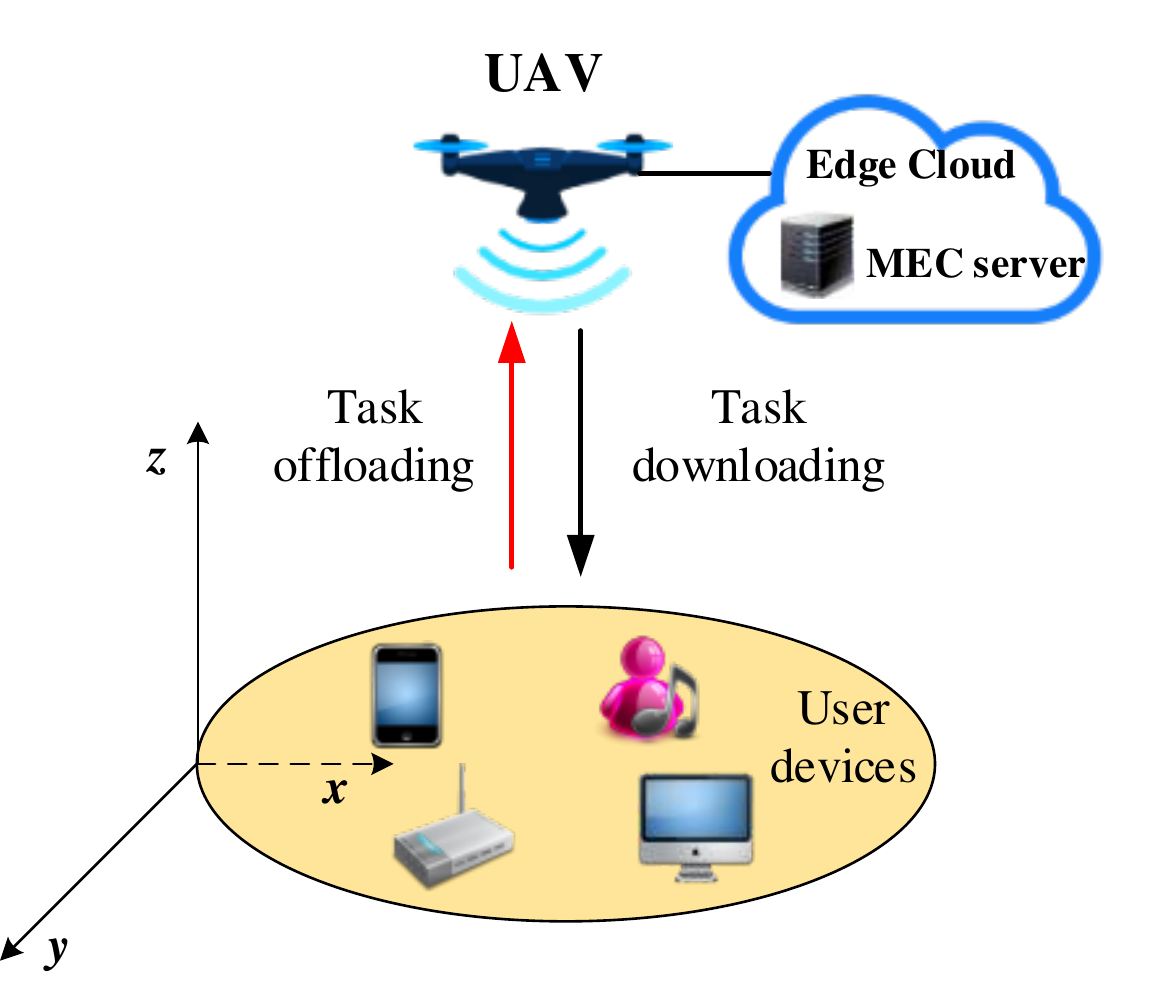}
\caption{Illustration of the UAV-mounted MEC system that provides application offloading opportunities to ground user devices.
}
\label{UAV-MEC}
\end{figure}

In UAV-enabled networks, the resource-constrained mobile devices are able to offload their computation-intensive tasks to a flying UAV with high
computing ability and flexible connectivity at the edge of network, thereby saving their energy and reducing traffic load at
the fixed cloud servers. Therefore, the UAV equipped with an MEC server offers promising advantages compared to the conventional ground cellular network with fixed BSs.
The system model of UAV-mounted MEC is demonstrated in Fig. \ref{UAV-MEC}.
In such a case, each mobile device needs to decide either local computing or edge computing. For the former, the mobile devices can locally execute
their own tasks by the embedded micro-processor, this will occupy their local computation resources and consume large quantities of energy. For the latter, the mobile devices are allowed to offload their intensive computation tasks to the MEC server co-located in UAVs directly, and then the MEC server will execute the computation tasks on behalf of the mobile devices. Actually, each mobile device is associated with a  nearby UAV node who currently has enough battery power and computing resources.
Table \ref{RMEC} shows a significant body of works on UAV-based MEC.

To expound a litter further,
the idea of installing the MEC processor on a UAV was initially putted forth by Jeong \textit{et al.} in \cite{Jeong2017IET} that offered the computation offloading opportunities to mobile devices. In this work, the authors considered a mobile device and a UAV with the aim of minimizing the energy consumption of mobile device by optimizing the bit allocation for uplink/downlink communication under the condition of a predetermined UAV's trajectory.
After that, Jeong \textit{et al.} \cite{Jeong2017mobile} presented extension to the multiple mobile devices  setting,
the problem of optimizing the bit allocation and UAV's trajectory was tackled  for minimizing the mobile energy consumption subject to the latency and UAV's energy budget constraints, where both the OMA and NOMA schemes were considered respectively.
Tang \textit{et al.} \cite{Tang2018novel} discussed a highly dynamic social network  supported by UAV-mounted cloudlets where UAVs were leveraged to quickly construct a meshed offloading backbone.
With the deployed UAVs with high performance computing cloudlets, both computation and traffic load could be shared from central
cloud to the edge of network, such that the computing burden of the cloud servers and the traffic load were reduced.
Zhou \textit{et al.} \cite{Zhou2018uav} concentrated their attention on the power minimization problem in a UAV-enabled MEC system by jointly optimizing the computation offloading and trajectory design. In this considered situation, UAV played dual-role namely, it not only performed the computation tasks offloaded from mobile devices, but also acted as a source transmitter to recharge the battery-powered devices.

On the other side, the on-board processor in a UAV may not have enough computing resources due to the small
size of UAV. This puts a restraint on efficient execution of complex applications. The battery and computation performance
of UAV can be virtually enhanced by leveraging computation offloading to remote cloud server or nearby edge servers via MBS or SBSs.
Therefore, a UAV node has an option to either process application using its own resources, or send its computation tasks to edge server or a remote cloud for processing according to applications' QoS requirements.
When UAVs deal with high resolution images quickly,
Jung \textit{et al.} \cite{Jung2017acods} addressed the problem of insufficient energy and computing resources via offloading the data processing to a GCS for prolonging UAV's flying time. After the GCS completed the image processing, the GCS returned the processed images to UAVs for the current operational mission.
According to a use case of face recognition, Motlagh \textit{et al.} \cite{Motlagh2017uav}  considered the offloading of video data processing from UAVs to an MEC node. Also, a testbed was developed from the viewpoint of a practical implementation to show the performance gains of the MEC-based offloading approach over the local processing of video data onboard UAVs in terms of energy consumption and processing time.
Additionally, the problem of energy consumption for computation
tasks offloading from multi-UAV to ground BS was pursued by Hua \textit{et al.} in \cite{Hua2018EnergyO}, where four types of access schemes in
the uplink transmission were proposed and compared.

\subsection{Caching in the Sky}

\begin{table*}
\centering
\setlength{\tabcolsep}{2pt}
\renewcommand{\arraystretch}{1.5}
\extrarowheight 3pt
\caption{Summary of Contributions to Caching in the Sky.}
\begin{tabular}{| l | l | l | l | l | }
\rowcolor{gray!15}
\hline
 \textbf{Reference} & \textbf{Objective} & \textbf{Mobility} & \textbf{Types of BSs} & \textbf{Number of UAVs} \\
\hline
\hline
Chen \textit{et al.} \cite{Chen2017liquid} & UAV cache & Static & UAV-only & Multiple UAVs \\
\hline
Chen \textit{et al.} \cite{Chen2017caching} & UAV cache & Mobile &  Hybrid (UAVs \& remote radio heads) & Multiple UAVs \\
\hline
Xu \textit{et al.} \cite{Xu2017overcoming} & UAV cache & Mobile &  UAV-only &  Single UAV \\
\hline
Zhao \textit{et al.} \cite{Zhao2018caching}  & UAV cache & Static  & Hybrid (UAVs \& ground BSs) &  Multiple UAVs \\
\hline
Fang \textit{et al.} \cite{Fang2018Context}  & UAV cache & Mobile  & UAV-only &  Multiple UAVs \\
\hline
\end{tabular}
\label{RCaching}
\end{table*}

Wireless data traffic has been increasing dramatically in recent years due to the proliferation of new mobile devices and various mobile applications. The driving forces behind this traffic growth have fundamentally shifted from the steady increasing in demand for connection-centric communications (such as smart phones and text messages) to the explosion of content-centric communications (such as video streaming and popular music).
Even though SBSs are densely deployed to accommodate the vast amount of traffic, a heavy burden will be imposed on the backhaul links.
In fact, the backhaul network can not deal with the explosive growth in mobile traffic.
One promising method is to intelligently caching some popular contents at the network edge (i.e., UAVs, relays, or D2D devices),
such that the demands from users for the same popular contents can be accommodated easily without duplicate transmissions via the backhaul links.

In general, mobile users are constantly moving, and thus a more flexible caching strategy is desired.
UAV as a flying BS can dynamically cache the popular contents, track the mobility patterns of wireless devices and then effectively serve them.
This not only significantly reduces the transmission latency, but also alleviates the traffic offloading on the backhaul especially during peak-load time.
In UAV-assisted edge caching, the contents can be directly cached in the UAV-BSs and then distributed to users, or cached
in the D2D devices and scheduled by the UAV-BSs \cite{Cheng2018air}. For the former one, the contents can be cached at UAV-BSs during the off-peak times. For the latter one, mobile users can cache the contents requested, and distribute such contents among nearby users following the scheduling of ground BSs or UAV-BSs. Such edge caching strategies can enhance the quality of experience (\gls{QoE}) of users while reducing the needed backhaul link capacity. In this respect, a number of contributions on caching at UAV have been done, as shown Table \ref{RCaching}.

More specifically, Chen \textit{et al.} \cite{Chen2017liquid} presented an idea of locally caching the popular content at flying UAVs in an LTE unlicensed band system that allowed them to directly transmit data providing services for the ground users, where a dynamic resource allocation algorithm based on machine learning  was proposed to  autonomously learn and determine which content to cache and how to allocate the licensed and unlicensed bands.
In \cite{Chen2017caching}, the problem of proactive deployment of cache-enabled UAVs in a cloud radio access network was explored to minimize the transmit power of the UAVs. In particular, a novel machine learning framework of conceptor-based echo state networks was proposed to effectively predict the content request distribution and  mobility pattern of each user. Also, the optimal locations of UAVs and the contents to cache at UAVs were derived.
In \cite{Xu2017overcoming}, Xu \textit{et al.} discussed a proactive caching scheme to prolong the UAV endurance where the UAV pro-actively delivered the files to a subset of selected ground nodes that cooperatively cached all the files. And then the files can be retrieved by each ground node either directly from its local cache or from its nearest neighbor via D2D communications. Simulation results showed the great potential of proactive caching in overcoming the limited endurance issue.
In the meantime, Zhao \textit{et al.} \cite{Zhao2018caching} examined the caching UAV-assisted secure transmission in a hyper-dense network where the video streaming was cached at both UAVs and SBSs simultaneously.
The interference alignment was exploited to eliminate the interference between ground users, and the idle SBSs
were further utilized to disrupt the potential eavesdropping by generating the jamming signals.
Considering the practical influence of propagation groups (i.e., LoS and non-LoS), Fang \textit{et al.} \cite{Fang2018Context} designed a joint scheme of UAVs' caching contents and service locations in a multi-UAV-aided network with the aim of achieving the tradeoff between user's service probability and transmission overhead, where the formulated optimization problem was modeled as a UAV caching game and the optimal solution could be obtained.

\section{Future Research Directions \label{f}}
In spite of the potentials combining UAV with 5G techniques, the research on UAV-assisted wireless networks is still in its infant stage and many open issues are in need of further investigation. In this section, we shed light on the new opportunities in emerging
network architecture and highlight interesting research topics for future directions.

\subsection{Energy Charging Efficiency}

Energy limitation is the bottleneck in any UAV communications scenario. As recent developments in battery technologies such as enhanced lithium-ion batteries and hydrogen fuel cells, energy harvesting is used to extend the flight times by utilizing green energy sources (such as solar energy). However, the efficiency of energy harvesting is relatively lower due to longer distance and random energy arrivals. To enhance the charging efficiency, novel energy delivering technologies such as energy beamforming through multi-antenna techniques and distributed multi-point WPT are of great interest.

\subsection{UAV-to-UAV and Satellite-to-UAV Communications}

To provide communication service to ground wireless devices over a significantly wide area, a swarm of UAVs construct a multihop network to help the devices send and pick up packets, each of which has a trajectory.
However, due to the high-speed mobility and the need to maintain the close communication links with ground users,
the link connection with the neighboring UAVs is disconnected frequently. In this case, all the traditional routing protocols cannot work well in FANETs.
Therefore, how to control the flight of the UAVs to realize a good service is a challenging direction.
In addition, when multiple UAVs collaborate, collision avoidance also become a significant development for UAV safe operation.
On the other hand, state-of-the-art satellite-to-UAV channel models lack detailed propagation effects. The exploitation of channel propagation models for satellite-to-UAV communications
is still in its infancy and remains a topic for future research.

\subsection{Interaction of Different Segments}

For the integrated space-air-ground network, a major issue is how to take advantage of innovative techniques
to ensure seamless integration among the space-based network, the air-based network and the ground cellular network.
Thus, it is desirable to design some cooperative incentives between different segments and dedicated cross-layer protocol designs are needed to ensure link reliability. In such a complex network environment, it is also important to provide scalable and flexible interfaces for these segments to interact and cooperate for achieving attractive benefits, i.e., how to implement the seamless information exchange and
data transmission among heterogeneous networks. For instance, the increasing variety of services may require UAVs to be the gateways between different networks, it is crucial in such a complex network to design interworking mechanisms for ensuring link reliability.

\subsection{Synergy of UAVs and IoT systems}
The Internet of UAVs (\gls{IoUAVs}) is a concept first introduced by Gharibi \textit{et al.} \cite{Gharibi2016IoD},
which argues the intersection of both existing IoT with UAVs in a dynamic integration.
Due to the unique characteristics, such as fast deployment, easy programmability, fully controllable mobility, and scalability,
IoUAVs are a promising solution to realize the framework of future IoT ecosystem where humans, UAVs, and IoT devices interact
on a cooperative basis, which enable ubiquitous information sharing and fine-granularity coordination among a fleet of UAVs.
In spite of the huge potential benefits of IoUAVs, the endurance and reliability performance is fundamentally limited by the maximum battery capacity, which is generally small due to practical SWAP constraints. On the other hand, additional energy consumption is required for IoUAVs to support mobility
and avoid collision, which is usually several orders of magnitudes higher than the energy consumed for data delivery, and relies on trajectory variations in the timescale of seconds especially in industrial IoUAVs \cite{Zhou2018Energy}.
Therefore, how to achieve an energy-aware synergy between the angles of UAVs and IoT systems is non-trivial.
Another worthwhile aspect is on
how to exploit the synergy between UAV mobility and user mobility for improving the efficiency and increasing the profitability of
wireless networks \cite{Yaliniz2018Spatial}.
Besides, the synergy of IoT and UAVs remains largely an untapped field of future technology that has the potential
to bring about drastic changes to how we live today.

\subsection{Security and Privacy}
The integrated network may face malicious attacks due to the open
links and dynamic topologies that blanket out a mission-critical area by intentional jamming/disruption.
In UAV-aided networks, the security is important since UAVs are always unattended, which leaves them easily captured or attacked. To avoid malicious modification, there is a need for a secure and lightweight mechanism to prevent attacks such as eavesdropping, man-in-the-middle attack, and so on.
Artificial intelligence solutions were proposed for addressing the security in cellular-connected UAV application use cases \cite{Challita2018Artificial}, while a zero-sum network interdiction game was advocated to capture the cyber-physical security threats
in UAV delivery systems \cite{Sanjab2018Prospect}. In the large coverage area of
space-air-ground integrated networks, SDN controllers are responsible for managing resources and controlling network operation,
it is urgent to protect the SDN controllers from different cyber-attacks where the adversaries are able to wiretap the data and control signals
transmitted through the radio links of the UAV systems.
The cyber-attacks to the UAV systems have been reported in \cite{Javaid2012cyber} and the cyber-security is
still a significant challenge to be overcome in the true utilization of UAVs. Therefore, designing timely strategies and counter-mechanisms are required to counteract malicious cyber-attacks.

\subsection{Space-Air-Ground Integrated Vehicular Networks }
Integrating space-air-ground communications into vehicular networks can provide high data rate for vehicular users  in urban/suburban areas by ground network, ubiquitous connectivity between vehicles in rural and remote areas by satellite network, as well as coverage expansion of infrastructures and network information collection in poor or congested areas by UAVs \cite{Zhang2017software}.
For this reason, the work \cite{Shi2018drone} proposed a UAV-assisted framework to integrate UAVs with
ground vehicular networks for efficiently augmenting the system performance.
In the ecosystem of space-air-ground communications, the high mobility of satellites and UAVs will change the propagation channel state all the time in terms of free space path loss and Doppler effect.
To cope with the interworking issues between space-air-ground networks and vehicular networks, effectively designed network architecture is required.  Going further, to support the data delivery with low latency and high reliability, a comprehensive control mechanism coordinating the spectrum allocation, link scheduling and protocol design for the space-air-ground
propagation channel needs to be further considered.

\subsection{Integration of Networking, Computing, and Caching}

Despite existing studies have been done on networking, computing, and  caching in wireless networks separately, the joint consideration of the three advanced techniques should be carefully designed in a systematic way to meet the intrinsic requirements of next generation smart IoT, and even make a trade-off between the operation costs (e.g., energy consumption) and performance benefits (e.g., decreasing latency).
Literature \cite{Huo2016Software} developed an architecture
for the integration of software defined networking, caching and computing, and detailed
the key components of data, control, and management planes.
Later, literature \cite{He2017Software} proposed a big data deep reinforcement learning approach to enable dynamic orchestration of networking, caching, and computing resources for improving the performance of applications in smart cities. Fully utilizing the networking, computing, and caching technologies can essentially complement the current
development of IoT, however, new features also create unexpected
problems that cannot be directly addressed through the traditional approaches designed for low-rate IoT systems. Thus, how
to effectively integrate existing capabilities to address the fundamental problems in smart IoT remains a topic for future research.

\subsection{Environment Uncertainty}

Since future wireless networks can provide heterogenous communication, computation, and caching resources \cite{Markakis2017Mag}, it is of great importance to efficiently utilize these heterogenous resources to support different big data applications.
Zhang \textit{et al.} \cite{Zhang2018Synergy} focused their attention on the synergistic
and complementary features of big data and 5G ecosystem that allowed service, content, and function providers to deploy their services/content/functions at the network edges, and the data network aided data acquisition and big data assisted edge content caching were provided.
Since massive network data can be utilized to train prediction models to predict future network events, the
proactive actions can be performed in advance to avoid network faults or service failures. For this purpose, accurate prediction, such as for spatial-temporal traffic distribution, service/content popularity, and user mobility, is required to facilitate optimal decision making and thus improves the overall network performance.

\subsection{Other Interesting Topics}

Apart from the above-discussed prospects, there are still many open issues related to the practicality of performing UAV communications.
For instance, in certain application scenarios (such as in forests), there may exist obstacles and rich scatterers between the UAV and ground users, thus a  more realistic air-to-ground channel model that incorporates temperature, wind, foliage, and urban environments is an interesting problem worth future research efforts. Furthermore, in UAV-enabled multi-user NOMA systems, it has been shown that the optimal user clustering and user-pairing algorithms are underexplored fields. Besides, new unmanned aircraft traffic management systems may be necessary to safely handle the high density of low altitude UAV traffic \cite{Yang2018Telecom}, which is responsible for the cooperative path planning and collision avoidance of multiple UAVs. The UAV-based antenna array system is another footprint for providing high data rate and low
service time \cite{Mozaffari2018Communications}, since the number of antenna elements (i.e., the number of UAVs) is not limited by space constraints.
To prevent privacy leakage of UAV communication and ensure the
integrity of collected data from UAVs, blockchain technology
(i.e., \textit{aerial blockchain}) is expected to be a new paradigm to securely and
adaptively maintain the privacy preferences during UAVs and GCS communication process.

\section{Conclusions \label{g}}
The number of mobile devices for IoT is growing rapidly, and there needs to be a high capacity and broadband connectivity communication system that can reliably support many IoT devices. To meet these requirements, the flying UAVs have attracted wide research interests recently.
In this survey, we provided a brief understanding on UAV communications in 5G/B5G wireless networks.
Particularly, we presented three major contributions: First, we have envisioned the space-air-ground integrated network for B5G communication systems. The related design challenges were discussed that can greatly help to better understand this newly introduced network architecture. Second, we have provided an overview of recent research activities on UAV communications combining the 5G techniques from the viewpoints of physical layer, network layer, and joint communication, computing and caching. In the end, we have unearthed several open research issues conceived for future research directions.
This is a timely and essential topic with the hope that it can serve as a good starting point for the IoT applications of 5G/B5G.

\bibliographystyle{IEEEtran}
\bibliography{refs}

\end{document}